\documentclass[%
 reprint,%
 amssymb, amsmath,%
 aip,cha,%
]{revtex4-1}
\usepackage{docs}%
\usepackage{dcolumn}
\usepackage{graphicx}
\usepackage{xcolor}
\usepackage{amsfonts}
\usepackage{amsmath}
\usepackage{amssymb}
\usepackage{amsthm}
\usepackage{bm}
\usepackage[colorlinks=true, linkcolor=blue, citecolor=blue, urlcolor=blue]{hyperref}

\date{\today}

\begin{document}

\title{Bifurcations and degenerate periodic points in a three dimensional chaotic fluid flow}

\author{L.~D. Smith}
 \email{lachlan.smith@monash.edu}
  \affiliation{ 
Department of Mechanical and Aerospace Engineering, Monash University, Clayton, VIC 3800, Australia
}%
 \affiliation{CSIRO Mineral Resources, Clayton, VIC 3800, Australia}
\author{M. Rudman}%
\affiliation{ 
Department of Mechanical and Aerospace Engineering, Monash University, Clayton, VIC 3800, Australia
}%
\author{D.~R. Lester}
\affiliation{School of Civil, Environmental and Chemical Engineering, RMIT University, Melbourne, VIC 3000, Australia}
\author{G. Metcalfe}
\affiliation{CSIRO Manufacturing, Highett, VIC 3190, Australia}
\affiliation{Department of Mechanical and Product Design Engineering, Swinburne University of Technology, Hawthorn, VIC 3122, Australia}
\affiliation{School of Mathematical Sciences, Monash University, Clayton, VIC 3800, Australia}

\begin{abstract}  
Analysis of the periodic points of a conservative periodic dynamical system uncovers the basic kinematic structure of the transport dynamics, and identifies regions of local stability or chaos. While elliptic and hyperbolic points typically govern such behaviour in 3D systems, degenerate (parabolic) points also play an important role. These points represent a bifurcation in local stability and Lagrangian topology. In this study we consider the ramifications of the two types of degenerate periodic points that occur in a model 3D fluid flow. (1) Period-tripling bifurcations occur when the local rotation angle associated with elliptic points is reversed, creating a reversal in the orientation of associated Lagrangian structures. Even though a single unstable point is created, the bifurcation in local stability has a large influence on local transport and the global arrangement of manifolds as the unstable degenerate point has three stable and three unstable directions, similar to hyperbolic points, and occurs at the intersection of three hyperbolic periodic lines. The presence of period-tripling bifurcation points indicates regions of both chaos and confinement, with the extent of each depending on the nature of the associated manifold intersections. (2) The second type of bifurcation occurs when periodic lines become tangent to local or global invariant surfaces. This bifurcation creates both saddle--centre bifurcations which can create both chaotic and stable regions, and period-doubling bifurcations which are a common route to chaos in 2D systems. We provide conditions for the occurrence of these tangent bifurcations in 3D conservative systems, as well as constraints on the possible types of tangent bifurcation that can occur based on topological considerations. [This paper is available from AIP Publishing at \url{http://dx.doi.org/10.1063/1.4950763}]
\end{abstract}


\maketitle

\begin{quotation}
Periodic points play a pivotal role in the organisation of transport and mixing in periodic fluid flows. Locally stable (elliptic) periodic points indicate regions of non-mixing, whereas  locally unstable (hyperbolic) periodic points are necessary for chaos and mixing. There is a third type of periodic point, known as a degenerate (parabolic) point, for which the only local deformation of fluid elements is shear. These points are often overlooked but also play an important role as they are on the brink of the stable/unstable classification and indicate abrupt changes in local stability. We demonstrate the impact of these degenerate points on transport in a model 3D fluid flow. In particular we discuss period-tripling bifurcations that occur when the rotation of elliptic points reverses direction, and tangent bifurcations that occur when periodic lines become tangent to invariant surfaces.
\end{quotation}

\section{Introduction}

Despite their ubiquity, the properties and organization of passive tracer transport in 3D time-periodic flows has received less attention than 2D flows\cite{Wiggins}, largely because the correspondence between 2D incompressible flows and one degree-of-freedom Hamiltonian systems breaks down for 3D systems. The correspondence between 2D incompressible flow and Hamiltonian mechanics means that for such systems, the tools and techniques of over a century of research into Hamiltonian chaos can be directly applied to understanding fluid mixing. Moreover, the additional spatial dimension results in an explosion of geometric complexity, creating more possibilities for transport and mixing structures. The combination of these factors means that transport and mixing in 3D incompressible flows is a rich source of ongoing research.

Central to the dynamical systems approach for conservative periodic systems is the analysis of periodic points, i.e. those that return to their initial position after some number of flow periods. In both 2D and 3D these provide the backbone of the kinematic template which governs transport and mixing. Elliptic points indicate a region of local stability and are generally associated with non-mixing regions, whereas hyperbolic points generate the stretching and folding motions which lead to chaos \cite{Ottino}. In 2D systems bifurcations of periodic points play a significant role in transport organization, corresponding to abrupt changes in the topology of coherent structures. These bifurcations are typically controlled by a single perturbation parameter. In contrast, periodic points in 3D flows may be found as periodic lines \cite{Pouransari,Mullowney+Julien+Meiss,Moharana2013}, a curve of periodic points, and the bifurcations exhibited by 2D systems can arise in 3D systems with the spatial dimension normal to locally 2D transport acting as the perturbation parameter. This is made clearer in 3D flows which admit a single invariant, such that transport is confined to a nested set of 2D invariant surfaces. In these cases the direction normal to the invariant surfaces can act as a perturbation parameter for the nested set of 2D systems \cite{Gomez2002}. 

Bifurcation points on periodic lines are necessarily of degenerate type (also known as parabolic type), where only shearing of local fluid can occur. While often overlooked, these points can have a significant effect on global transport \cite{Pouransari}, representing points of bifurcation in local stability and transport topology.

To probe the influence of degenerate points on transport and mixing, we study the bifurcations that occur on periodic lines in a 3D model flow, the 3D Reoriented Potential Mixing (3DRPM) flow, that is driven by a periodically reoriented dipole flow. We demonstrate a type of period-tripling bifurcation that is seen in some 2D systems \cite{Berry1977,Dullin2000,Barrio2010} and has been observed in a 3D model flow \cite{Mullowney+Julien+Meiss}, but its implications for transport have not been fully explored. In 2D systems the period-tripling bifurcation occurs as the merging of three period-3 hyperbolic points at a degenerate period-1 point. Unlike period-doubling bifurcations where the period-2 points appear after the bifurcation, the period-3 points exist both before and after the period-tripling bifurcation, such that there is only a single value of the bifurcation parameter where the period-3 points do not exist. It is for this reason the bifurcation has been described as a `touch-and-go' bifurcation \cite{Barrio2010}. In 2D this bifurcation results in a reversal in orientation of local Lagrangian structures, but the bifurcation itself has little impact on global transport properties. On the other hand, for 3D systems the third spatial direction can act as the bifurcation parameter for locally 2D transport, and period-tripling bifurcations occur where three period-3 hyperbolic lines intersect a period-1 line. These period-1 and period-3 lines are extensive and their manifolds are even more extensive, forming the kinematic template for a large portion of the flow domain. This entire transport structure is organized by the single bifurcation point, and is observed to destroy invariant tori, create `sticky' regions where particles are loosely trapped, create chaotic regions and affects transport in a region of the domain that is vastly more extensive that just the `neighborhood' of the bifurcation point.

In standard periodic point analysis, the eigenvalues of the Jacobian determine whether periodic points are elliptic, hyperbolic or degenerate. The corresponding eigenvectors have been used to determine the direction of continuation of periodic lines as well as the directions of contraction and expansion. We demonstrate an additional property of the eigenvectors in 3D conservative systems, showing that points where the eigenvectors become linearly dependent (coplanar) correspond to bifurcation points. The bifurcations at these points are constrained by the `Poincar\'{e} index', a conserved topological quantity. We detail two of the possible bifurcations that can occur: saddle--centre and period-doubling bifurcations, with emphasis on the transport barriers and chaotic regions that can be created by saddle--centre bifurcations. In systems which admit an invariant, these types of bifurcations occur when the periodic line becomes tangent to the invariant surfaces, providing a simple tool for their detection and analysis. In the absence of an invariant, such as in the 3DRPM flow, the system still admits local invariants based on local 2D approximation and these bifurcations occur at points where the periodic line becomes tangent to the iso-surfaces of the local invariant. We call this type of bifurcation a tangent bifurcation \cite{Note1}, since they occur at points where periodic lines meet local/global invariants tangentially.

\section{The 3DRPM flow}

As a model for transport and mixing in 3D fluid flows, in particular porous media flows, we consider a periodically reoriented 3D dipole flow, the 3D Reoriented Potential Mixing (3DRPM) flow\cite{Lester2012,Smith2012,Smith2014}. It is a natural three-dimensional extension of the 2D RPM flow which has been studied in the context of contaminant remediation and heat extraction/injection in groundwater flows \cite{Lester,Metcalfe2,Trefry,Sheldon2015}. A sink/source pair is used to model the extraction/injection of fluid produced by valved well-bores used in groundwater applications\cite{Mays2012}.

\subsection{Steady dipole flow}

\begin{figure}
\includegraphics[width=0.9\columnwidth]{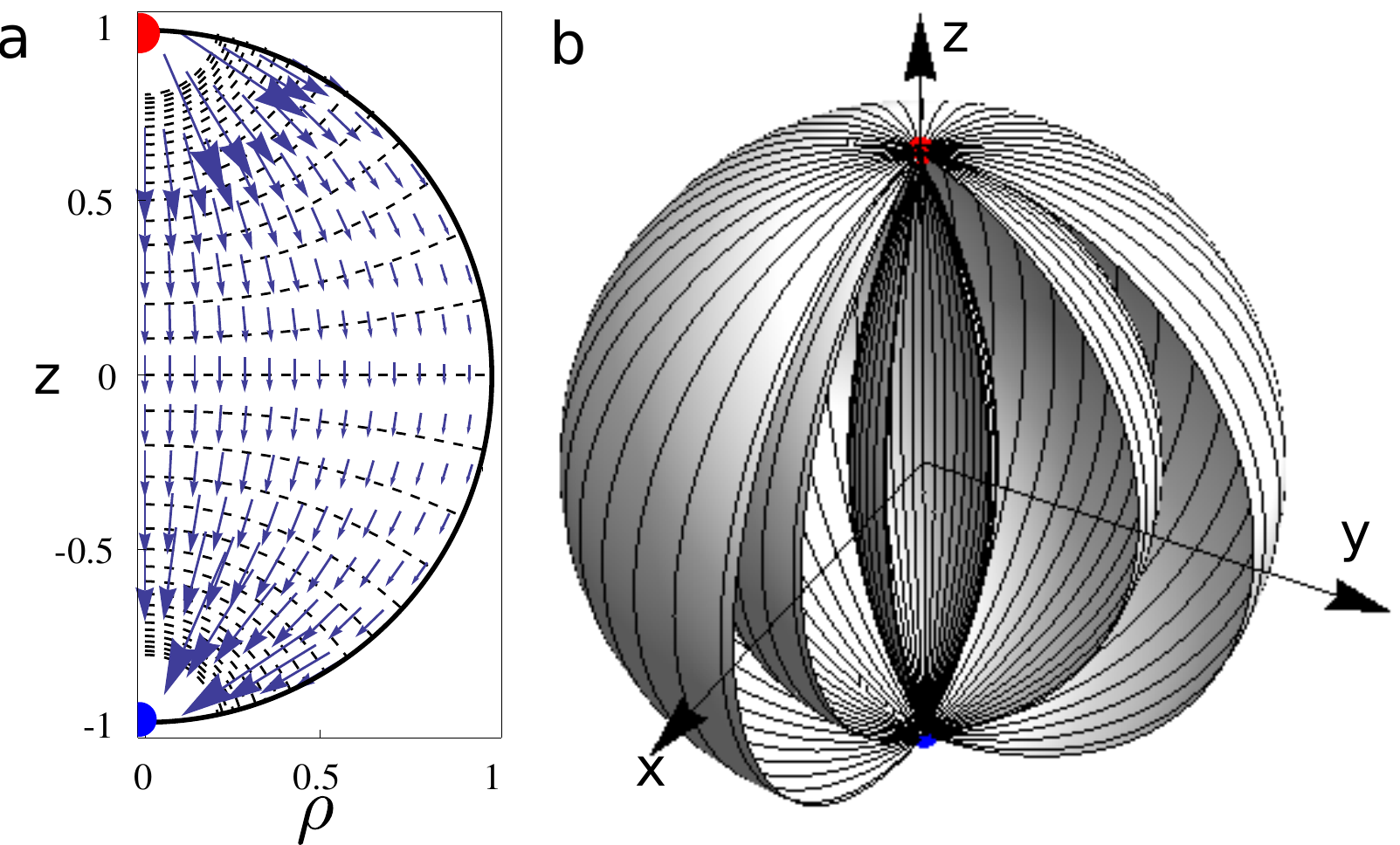}
\caption{Steady dipole flow. (a)~Contours of the axisymmetric potential function $\Phi$ and velocity field $\bm{v}$. (b)~Isosurfaces of the axisymmetric stream function $\Psi$.}
\label{fig:stream and potential}
\end{figure}

A steady dipole flow forms the basis for the time-dependent reoriented flow. It is driven by a source/sink pair located at $\bm{z}^\pm=(0,0,\pm 1)$. While the open dipole flow in an infinite 2D domain possesses a separating streamline coinciding with the unit circle, the corresponding stream-surface in 3D does not coincide with the unit sphere. We therefore confine the flow to the unit sphere $\Omega$, using a free-slip boundary condition. This flow is axisymmetric about the $z$-axis, admitting an axisymmetric Stokes stream-function $\Psi$, such that $\bm{v} = \nabla \times (\Psi/\rho)\hat{\bm{e}}_\theta$ where $(\rho,\theta,z)$ denote cylindrical coordinates. Combining incompressibility with the boundary conditions yields the equations governing the flow potential $\Phi$, where $\bm{v}=\nabla\Phi$,
\begin{equation}
\label{eq:potential_eqns}
\nabla^2 \Phi = 0, \,\, \text{and}\,\, \, \bm{n}\cdot \nabla \Phi \big|_{\partial \Omega} = \delta(z-1) - \delta(z+1),
\end{equation}
where $\bm{n}$ is the outward normal to the boundary $\partial \Omega$ and $\delta$ is the Dirac delta function. We solved these equations using the method of images \cite{Dassios+Sten} to find an analytic expression for the flow potential $\Phi$. In turn analytic expressions can be found for the velocity field and the axisymmetric Stokes stream-function $\Psi$:
\begin{align}
\Phi(\rho,\theta,z) &= \frac{1}{4 \pi }\left( \frac{2}{d^-} -\frac{2}{d^+} + \log \left(\frac{d^+-z+1}{d^-+z+1}\right) \right) \\
\Psi(\rho,\theta,z) &= \frac{1-\rho ^2-z^2}{4 \pi } \left(\frac{1}{d^-}+\frac{1}{d^+}\right)
\end{align}
where $d^\pm = \sqrt{\rho^2+(z\mp 1)^2}$ are the distances from the poles $\bm{z}^\pm$. Contours of $\Phi$ and $\Psi$ are shown in Fig.~\ref{fig:stream and potential} together with the velocity field $\bm{v}$. Both the stream-function $\Psi$ and azimuthal angle $\theta$ are invariants of the steady flow, and fluid particles follow streamlines given by intersections of these two isosurfaces. These streamlines are illustrated in Fig.~\ref{fig:stream and potential}b as solid lines on the surfaces.

As the steady dipole flow is a potential flow, it is also a Darcy flow \cite{Homsy1987, Metcalfe1}, i.e. it can be expressed in the form
\begin{equation}
\bm{v}(\bm{x}) = -\frac{K}{\mu} \nabla P(\bm{x})
\end{equation}
where $\mu$ is the fluid viscosity, $K$ is the permeability of the porous media, and the pressure gradient $\nabla P$ is obtained by scaling the flow potential $\Phi$. Therefore, the steady dipole flow also serves as a model for homogeneous porous media flow, with the dipole mimicking the action of injection and extraction of fluid.

In this study we create a closed flow by enforcing a reinjection protocol at the source/sink. We specify that particles that reach the sink are immediately reinjected at the source along the same streamline. This is an arbitrary choice and other valid reinjection protocols exist (see Lester et al.\cite{Lester} for examples of several choices) but this reinjection choice has the advantage of preserving Lagrangian structures during the reinjection process.

Here we denote by $\hat{Y}_t$ the solution of the advection equation
\begin{equation}
\label{eq:adv_eqn}
\dot{\bm{x}} = \bm{v}(\bm{x},t)
\end{equation} 
in the Lagrangian frame, describing streamlines as functions of time from an initial condition $\bm{X}$. The map $\hat{Y}_t$ satisfies
\begin{equation}
\label{eq:ds_flow}
\hat{Y}_0 (\bm{X}) = \bm{X}, \,\, \text{and} \,\, \frac{d}{dt}\hat{Y}_t(\bm{X}) = \bm{v}\big(\hat{Y}_t(\bm{X})\big) 
\end{equation} 
where $\bm{X}$ denotes Lagrangian coordinates. Since the velocity $\bm{v}$ is incompressible it follows that the advection map $\hat{Y}_t$ is volume-preserving for each value of $t$. We use the explicitly volume-preserving integration method of Finn and Chac\'{o}n\cite{Finn+Chacon} to numerically solve the advection equation (\ref{eq:adv_eqn}).

\subsection{The time-dependent flow}

\begin{figure}[b]
\includegraphics[width=0.9\columnwidth]{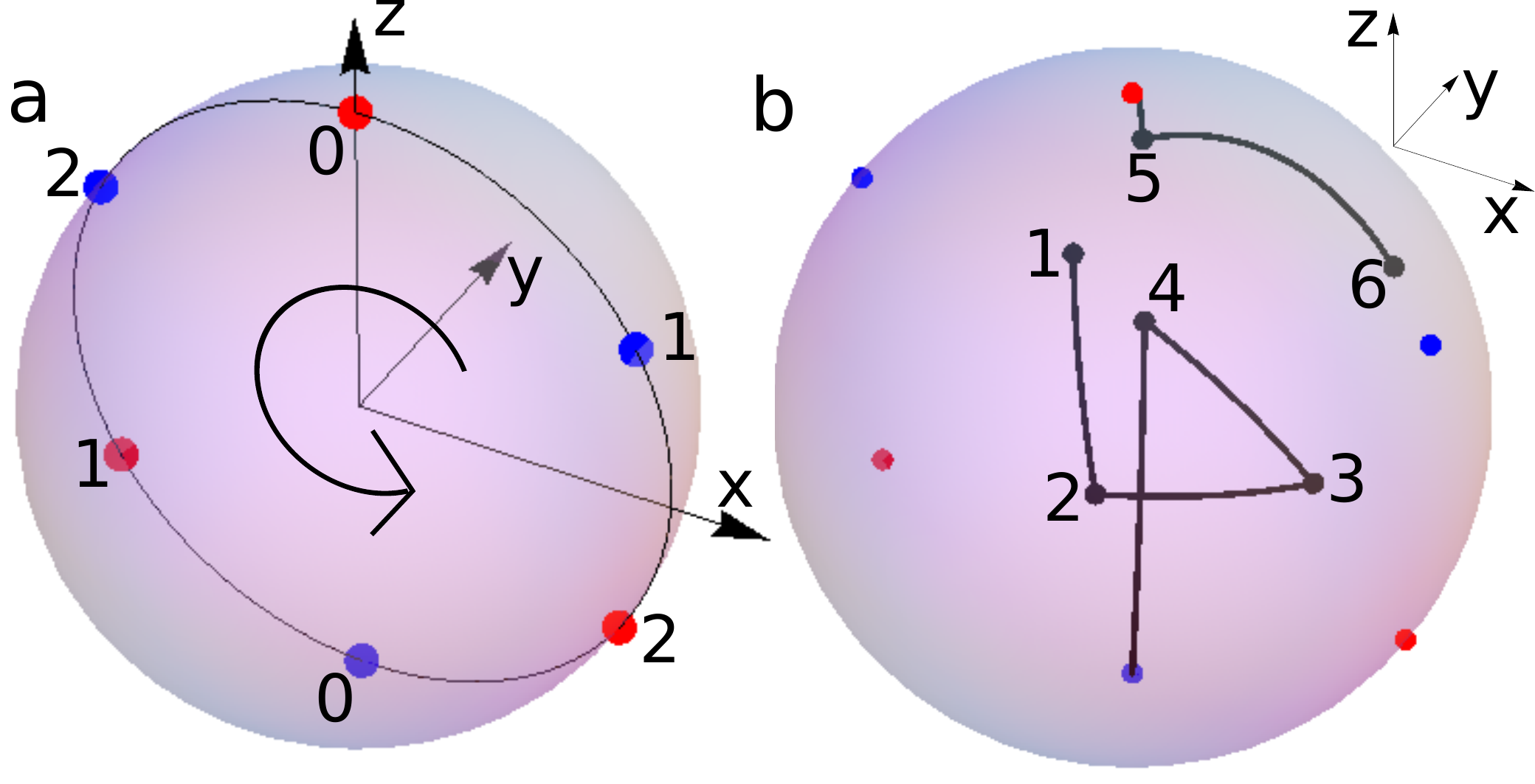}
\caption{The 3DRPM flow. (a)~Reorientation protocol for $\Theta=2\pi/3$. Dipole pairs are labelled according to the number of reorientations of the base flow modulo $3$. (b)~A typical particle trajectory for the protocol $(\tau,\,\Theta)=(0.3,\,2\pi/3)$.}
\label{fig:transient}
\end{figure}

In the steady dipole flow passive particles are confined to streamlines of constant azimuth $\theta$ and constant Stokes stream-function $\Psi$, and thus the flow cannot become chaotic. To create the crossings of streamlines required for chaotic motion we create a time-dependent flow by periodically reorienting the dipole. This allows fluid stretching at the dipoles to persist given appropriate flow parameters. Here we only consider the simplest reorientation protocol involving rotation of the dipole about the $y$-axis, providing the closest resemblance to the 2D RPM flow. The dipole is switched on for a time period $\tau$, then switched off, instantaneously rotated by $\Theta$ about the $y$-axis, and switched back on. We non-dimensionalize the reorientation period $\tau$ such that $\tau=1$ corresponds to the emptying time of the sphere under the steady dipole flow, i.e. the time it takes for all fluid in the sphere to pass through the sink. In this study we exclusively use a rotation angle of $\Theta=2\pi/3$, the dipole positions for this case are shown in Fig.~\ref{fig:transient} by the blue (sink) and red (source) points. The results in this study are generic to all rotation angles of the form $2\pi m /n$ with $n$ odd, except period-$n$ structures replace the period-3 structures that we observe.

The velocity field in the time-dependent flow can be approximated\cite{Lester2012,Smith2012} by the inertialess piecewise-steady velocity
\begin{equation}
\label{eq:piecewise_steady_approx}
\hat{\bm{v}}(\bm{x},t) = \bm{v}\left(R_{\lfloor \frac{t}{\tau} \rfloor \Theta}^y \bm{x}, t\right)
\end{equation}
where $R_\beta^y$ is the rotation matrix corresponding to rotation through the angle $\beta$ about the $y$-axis and $\lfloor a \rfloor$ is the largest integer less than $a$. 

For convenience we track particles in the rotating dipole frame. Rather than rotating the dipole at the end of each reorientation period $\tau$, we counter-rotate the particles about the $y$-axis. Each step of the advection-reorientation cycle can therefore be expressed as
\begin{equation}
Y_\tau^{-\Theta} (\bm{x}) = R_{-\Theta}^y \hat{Y}_\tau (\bm{x}).
\end{equation}
This map is the main object of our study, and has identical Lagrangian dynamics to the 3DRPM flow in the (fixed) Eulerian frame. 

\subsection{Symmetries}

Flow symmetries play an important role in the overall organisation of transport structures, including those associated with periodic points and lines. Symmetries of the time-dependent flow can be derived from underlying symmetries of the steady flow and the reorientation protocol.

The steady dipole flow possesses two basic symmetries: axisymmetry about the $z$-axis and a reflection reversal symmetry in the $xy$-plane. Algebraically these can be written respectively as
\begin{align} \label{eq:steady_sym1}
&\hat{Y}_t = R_\gamma^z \hat{Y}_t R_{-\gamma}^z,\\
 \label{eq:steady_sym2}
&\hat{Y}_t = S_{xy} \hat{Y}_t^{-1} S_{xy}^{-1},
\end{align}
where $S_{xy}$ is reflection in the $xy$-plane. These symmetries yield three symmetries for the map $Y_\tau^\Theta$. First, as a special case of the axisymmetry property (\ref{eq:steady_sym1}), the dipole flow is symmetric in the $xz$-plane, i.e. 
\begin{equation}
\hat{Y}_t = S_{xz}\hat{Y}_t S_{xz}^{-1}.
\end{equation}
As a result $Y_\tau^\Theta$ satisfies
\begin{align}
Y_\tau^\Theta &= R_\Theta^y \hat{Y}_\tau = R_\Theta^y S_{xz}\hat{Y}_\tau S_{xz}^{-1} \nonumber \\ 
&= S_{xz}  R_\Theta^y \hat{Y}_\tau S_{xz}^{-1} =  S_{xz} Y_\tau^\Theta S_{xz}^{-1}
\end{align}
and is therefore also symmetric in the $xz$-plane. The $xz$-plane $P_{xz}$ is an invariant surface of the map $Y_\tau^\Theta$, i.e. $Y_\tau^\Theta(P_{xz})=P_{xz}$, therefore this symmetry guarantees that the dynamics in the $y^+$ and $y^-$ hemispheres mirror each other. As the $xz$-plane acts as an impenetrable barrier, dividing $\Omega$ in two, we need only consider transport in the $y^+$ hemisphere.

Similarly, as another case of the axisymmetry property (\ref{eq:steady_sym1}) the steady dipole flow is symmetric about the $yz$-plane
\begin{equation}
\hat{Y}_t = S_{yz}\hat{Y}_t S_{yz}.
\end{equation}
This translates to the map $Y_\tau^\Theta$ as
\begin{align}
Y_\tau^\Theta &= R_\Theta^y \hat{Y}_\tau = R_\Theta^y S_{yz}\hat{Y}_\tau S_{yz} \nonumber \\
 &= S_{yz}  R_{-\Theta}^y \hat{Y}_\tau S_{yz} =  S_{yz} Y_\tau^{-\Theta} S_{yz}
\end{align}
since $R_\Theta^y S_{yz} = S_{yz}  R_{-\Theta}^y$. Therefore changing $\Theta$ to $-\Theta$ results in a reflection through the $yz$-plane, but does not alter transport dynamics. This is the 3D extension of the 2D result that for periodically reoriented flows where the base flow has two symmetries that $+\Theta$ and $-\Theta$ are equivalent \cite{Metcalfe2010}.

The map $Y_\tau^\Theta$ also possesses a reflection reversal symmetry, inherited from the reflection reversal symmetry of the dipole flow (\ref{eq:steady_sym2}) as follows,
\begin{align} \label{eq:ref-rev_sym}
Y_\tau^\Theta &= R_\Theta^y \hat{Y}_\tau = R_\Theta^y S_{xy}\hat{Y}_\tau S_{xy} \\ \nonumber
& = R_\Theta^y S_{xy} (Y_\tau^\Theta)^{-1} R_\Theta^y S_{xy} = S_1 (Y_\tau^\Theta)^{-1} S_1
\end{align}
where $S_1= R_\Theta^y S_{xy}$. Direct computation shows that $S_1$ is the map that reflects a point through the plane $z=\tan(-\Theta/2)x$. Therefore structures in the orbit topology must also evolve symmetrically about this plane. This has a significant impact on the locations of periodic points, since it constrains all chains of periodic points to be distributed symmetrically about the symmetry plane.

\subsection{Approximately 2D transport} \label{sec:3DRPM_transport}

\begin{figure*}
  \begin{minipage}[c]{0.65\textwidth}
    \includegraphics[width=\textwidth]{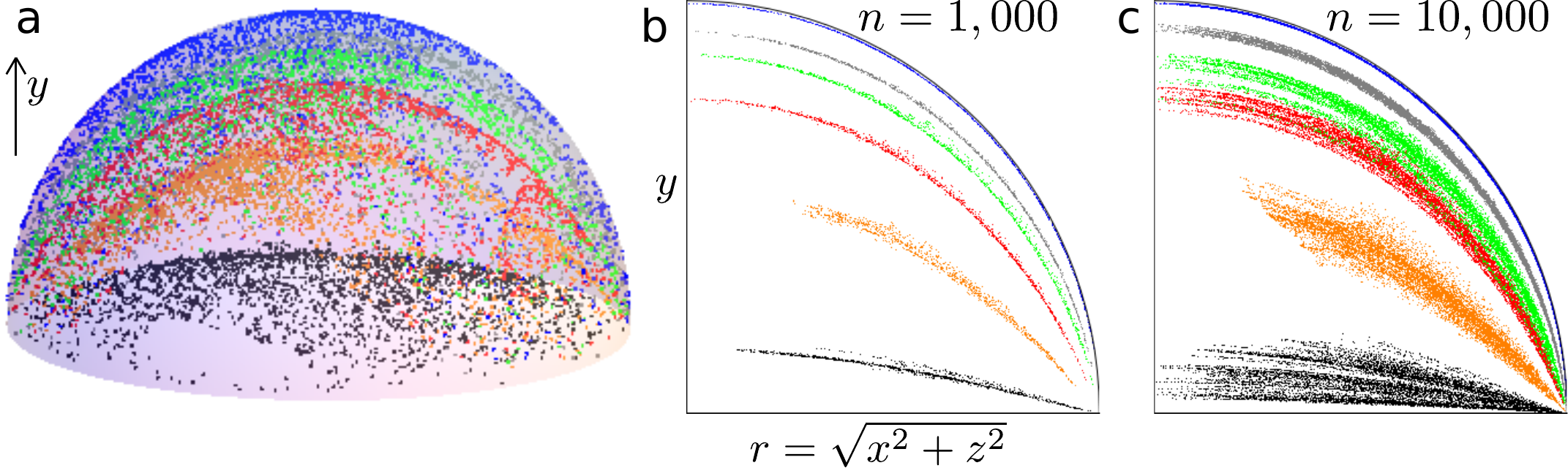}
  \end{minipage}\hfill
  \begin{minipage}[c]{0.3\textwidth}
    \caption{
       Poincar\'{e} sections in the 3DRPM flow with $\tau=0.328\approx 1.12 \tau_0$ ($\tau_0$ is defined in the text) and $\Theta=2\pi/3$. Colors correspond to six different initial particle locations, corresponding to different colors. (a) 3D view in the $y^+$ hemisphere after $2,000$ iterations. (b,c) Azimuthal projections of the Poincar\'{e} section into $(\sqrt{x^2 + z^2},y)$ coordinates after (b) $1,000$ and (c) $10,000$ iterations.
    } \label{fig:shells}
  \end{minipage}
\end{figure*}


Based on the similarity between the 3DRPM flow and the 2D RPM flow \cite{Lester,Metcalfe2,Trefry} and the presence of two invariant surfaces - the $xz$-plane and the spherical boundary - one might expect that particle transport in general would be confined to 2D surfaces. However, the 3DRPM flow does not admit a global invariant. At low values of $\tau$ (including those considered in this paper) particle motion is approximately 2D, as shown in Fig.~\ref{fig:shells}a,b by the approximately hemispherical shape of the Poincar\'{e} sections. However, for all values of $\tau$ there is transport transverse to these surfaces creating 3D motion, as is clearly illustrated in Fig.~\ref{fig:shells}c. This transverse transport has implications for the organization of coherent structures that we explore.

\section{Periodic points and lines}

Periodic points play a central role in the organization of fluid transport in periodic flows, determining mixing and non-mixing regions. A period-$n$ point of a map $\Lambda$ satisfies $\Lambda^n(\bm{x})=\bm{x}$, and the local stability near the periodic point is determined by the eigenvalues of the Jacobian
\begin{equation}
D\Lambda = \left( \frac{\partial \Lambda_i}{\partial x_j} \right)
\end{equation}
evaluated at the periodic point \cite{Ottino}. For a volume-preserving map, the product of the eigenvalues, $\lambda_1 \lambda_2 \lambda_3$ is equal to 1. If one of the eigenvalues, say $\lambda_3$, is equal to 1, and the others are not, then the Jacobian is diagonalizable and in an infinitesimally small region near the periodic point there is no transport in the direction of the corresponding eigenvector $\bm{v}_3$, called the {\em null direction}. Therefore, in this infinitesimally small region the map $\bm{x}' = Y_\Theta^\tau(\bm{x})$ can be written as
\begin{align} \label{eq:ess_2d}
\xi_1' &= f_1(\xi_1,\xi_2;\xi_3), \nonumber \\ 
\xi_2' &= f_2(\xi_1,\xi_2;\xi_3), \\
\xi_3' &=\xi_3, \nonumber
\end{align}
where $\xi_i$ corresponds to the coordinate in the direction of the eigenvector $\bm{v}_i$ \cite{Malyuga2002}, resulting in a 2D system contained in the plane spanned by the two transverse directions $\bm{v}_{1,2}$. In this case the periodic point forms part of a continuous line of periodic points, with the line continuing in the null direction. At each point on the periodic line the null direction provides a local invariant (valid in an infinitesimally small region) according to equation~(\ref{eq:ess_2d}); and conversely if there exists a local invariant in a region containing a periodic point then there is no transport transverse to the local invariant surfaces, providing a null direction. While it is sufficient, it is not necessary for a system to admit a global invariant for there to exist periodic lines, as we see for the 3DRPM flow.

There are three possible cases for the eigenvalues if there is a null direction. First if $\lambda_1=1/\lambda_2$ are both real, then the periodic point is locally unstable and called hyperbolic, fluid contracts along the direction corresponding to $\lambda_i<1$ and expands along the direction corresponding to $\lambda_i>1$. In 2D the stable manifold associated with a hyperbolic point $\bm{x}_0$, $W^s(\bm{x}_0)$, consists of all points that converge to $\bm{x}_0$ as the number of iterations of the map $\Lambda$ approaches infinity, i.e.
\begin{equation}
W^s(\bm{x}_0) = \left\lbrace \bm{x} \, : \, \lim_{N\to \infty} \Lambda^{N} \left(\bm{x}\right) = \bm{x}_0 \right\rbrace.
\end{equation}
Similarly, the unstable manifold consists of points that converge under the map $\Lambda^{-1}$, 
\begin{equation}
W^u(\bm{x}_0) = \left\lbrace \bm{x} \, : \, \lim_{N\to \infty} \Lambda^{-N} \left(\bm{x}\right) = \bm{x}_0 \right\rbrace.
\end{equation} 
These structures naturally extend to 3D systems with periodic lines since they can be thought of as a nested set of locally 2D systems according to equation~(\ref{eq:ess_2d}). The 1D manifolds of the nested 2D systems combine to create 2D manifold surfaces \cite{Ottino}
\begin{equation}
W^{s,u}_{2D} = \bigsqcup_{\bm{x}_0 \in \mathcal{S}} W^{s,u}(\bm{x}_0) 
\end{equation}
where $\sqcup$ is the disjoint union and $\mathcal{S}$ is a hyperbolic segment of a periodic line. An example of this for the 3DRPM flow is shown in Fig.~\ref{fig:saddle-node_manifolds} (multimedia view) where each curve (white to dark red) is the 1D unstable manifold for a single point $W^u(\bm{x}_0)$ on the hyperbolic segment of a period-3 line, the union of which form the 2D unstable manifold surface. Similar to 1D manifolds in 2D systems, the manifolds $W^{s,u}_{2D}$ are invariant, and form barriers to fluid transport as they are co-dimension 1. The geometry of these manifolds, and their intersections, plays a significant role for the overall transport dynamics. If the intersecting manifolds belong to the same hyperbolic point, then it is called a homoclinic connection, otherwise it is called a heteroclinic connection. If the intersection is transverse then they must intersect infinitely many times, creating the stretching and folding of fluid elements necessary for chaos. Conversely, tangent intersections form transport barriers that can confine fluid. It is possible that a single 2D manifold will have a combination of transverse, tangent, homoclinic and heteroclinic connections, as is the case in Fig.~\ref{fig:saddle-node_manifolds} (multimedia view), where it is seen that for greater $y$-values (closer to white) the 1D curves loop back to their initial position, forming tangent homoclinic connections, however at a critical value of $y$ the 1D manifolds develop a wavy pattern that indicates the presence of transverse connections.  

The second type of periodic point are those that are locally stable, called elliptic points. In this case particles are rotated about the periodic point, and the non-null eigenvalues of the Jacobian form a complex conjugate pair with $\lambda_1 = \bar{\lambda}_2$ and $|\lambda_1|=|\lambda_2|=1$. They must therefore be of the form $\lambda_{1,2} = \cos \alpha \pm i \sin \alpha$ where $\alpha$ is the angle of local rotation about the periodic point. Since there is a null direction, the local rotation occurs in the plane spanned by the vectors $\Re(\bm{v}_1),\,\Im(\bm{v}_1)$, where $\Re$ and $\Im$ denote the real and imaginary parts respectively. Enclosing elliptic segments of periodic lines is an invariant tube that creates a non-mixing region that is topologically distinct to mixing regions, i.e. particles can neither enter nor escape the tube.

The last type of periodic point -- called degenerate or parabolic -- is locally unstable with the eigenvalues of the Jacobian all equal to $\pm 1$. In this case there is local shearing of fluid, but no net rotation or stretching. These points can come in a number of different forms, resulting in different types of bifurcations of local transport dynamics.

When analysing periodic points in 2D systems, a useful quantity is the `Poincar\'{e} index' \cite{Katok}, which relates the number and type of periodic points to the topology of the manifold within which they are hosted. This index can be computed by forming a closed curve around the periodic point and calculating the number of counter-clockwise rotations of the velocity vector in one counter-clockwise traverse of the loop. Under this measure, hyperbolic points have Poincar\'{e} index $-1$ and elliptic points have Poincar\'{e} index $+1$. The sum of Poincar\'{e} indices is preserved under continuous deformation of a flow, and furthermore is related to the topological genus of the flow domain as per the Poincar\'{e}--Hopf theorem. This provides a constraint on the number and type of periodic points created or annihilated during bifurcation. When considering subdomains of the flow domain, the sum of the Poincar\'{e} indices within the subdomain also remains constant under continuous deformations of the flow, unless a periodic point crosses the boundary of the subdomain. The Poincar\'{e} index can also be applied to 3D systems with periodic lines, since the existence of a null direction means the system is essentially 2D in an infinitesimal region. Therefore the Poincar\'{e} index of the essentially 2D systems must remain constant as the periodic line is traversed, which can be loosely thought of as a requirement of topological continuity.

We use a numerical approach to find and classify periodic points in the 3DRPM flow, with the reorientation angle $\Theta=2\pi/3$, the focus is on period-1 and period-3 points since they have a dominant influence on the resulting Lagrangian dynamics. For rotation angles of the form $\Theta=2\pi m/n$ with $n$ odd, the period-1 and period-$n$ points will play a dominant role, while for even $n$ or rotation angles $\Theta$ that are incommensurate with $\pi$ (i.e. $\Theta/\pi$ is irrational), fundamentally different phenomena occur due to the difference in the nature of the degenerate point at the origin of the flow in the limit as $\tau \to 0$ \cite{Lester}. The reflection reversal symmetry (\ref{eq:ref-rev_sym}) means all period-1 points must be distributed symmetrically about the symmetry plane $z=\tan(-\Theta/2)x$. Furthermore, as the map $Y_\tau^\Theta$ is the composition of a map $\hat{Y}_\tau$ that preserves the azimuth $\theta=\arctan(y/x)$, and a rotation $R^y_\Theta$, it follows that period-1 points satisfy $\bm{x} = (Y_\tau^\Theta)^{-1}(\bm{x}) = (\hat{Y}_\tau)^{-1}R^y_{-\Theta} (\bm{x})$, and the azimuthal angle of $R^y_{-\Theta}(\bm{x})$ must equal the azimuthal angle of $\bm{x}=(x,y,z)$. This can be expressed as
\begin{equation}
\arctan\left(\frac{y}{x\cos \Theta - z \sin\Theta} \right) = \arctan \left( \frac{y}{x} \right),
\end{equation}
and implies that either $z=\tan(-\Theta/2)x$ (i.e. the point is on the symmetry plane) or $y=0$ and the point is on the $xz$-plane. Since the $xz$-plane is invariant and contains all the dipole locations, it is qualitatively similar to the 2D RPM flow, for which it can be shown that period-1 points must lie on the symmetry plane by essentially the same argument as above except using the streamfunction $\Psi$ instead of the azimuthal angle $\theta$ \cite{Lester}. This also applies to the 3DRPM flow, as the streamfunction $\Psi$ is conserved by the steady dipole advection $\hat{Y}_\tau$, and so a period-1 point must satisfy $\Psi(R^y_{-\Theta} \bm{x}) = \Psi(\bm{x})$. Therefore, all period-1 points must lie on the symmetry plane $z=\tan(-\Theta/2)x$. We also observe that all periodic points in the 3DRPM flow with period occur as smooth curves, each intersecting the $xz$-plane, i.e. we have not found any isolated odd periodic points. Therefore we may restrict our initial search for period-1 points to the line given by the intersection of the symmetry plane and the $xz$-plane, which significantly reduces the search space. Once solutions to $(Y_\tau^\Theta)^N (\bm{x}) = \bm{x}$ have been found, we compute the Jacobian to find the null direction and search for new periodic points a fixed distance $\delta$ away in the neighbourhood of the null direction. Once the full line has been found we again use the Jacobian to classify periodic points using the associated eigenvalues.

\begin{figure}
\includegraphics[width=0.8\columnwidth]{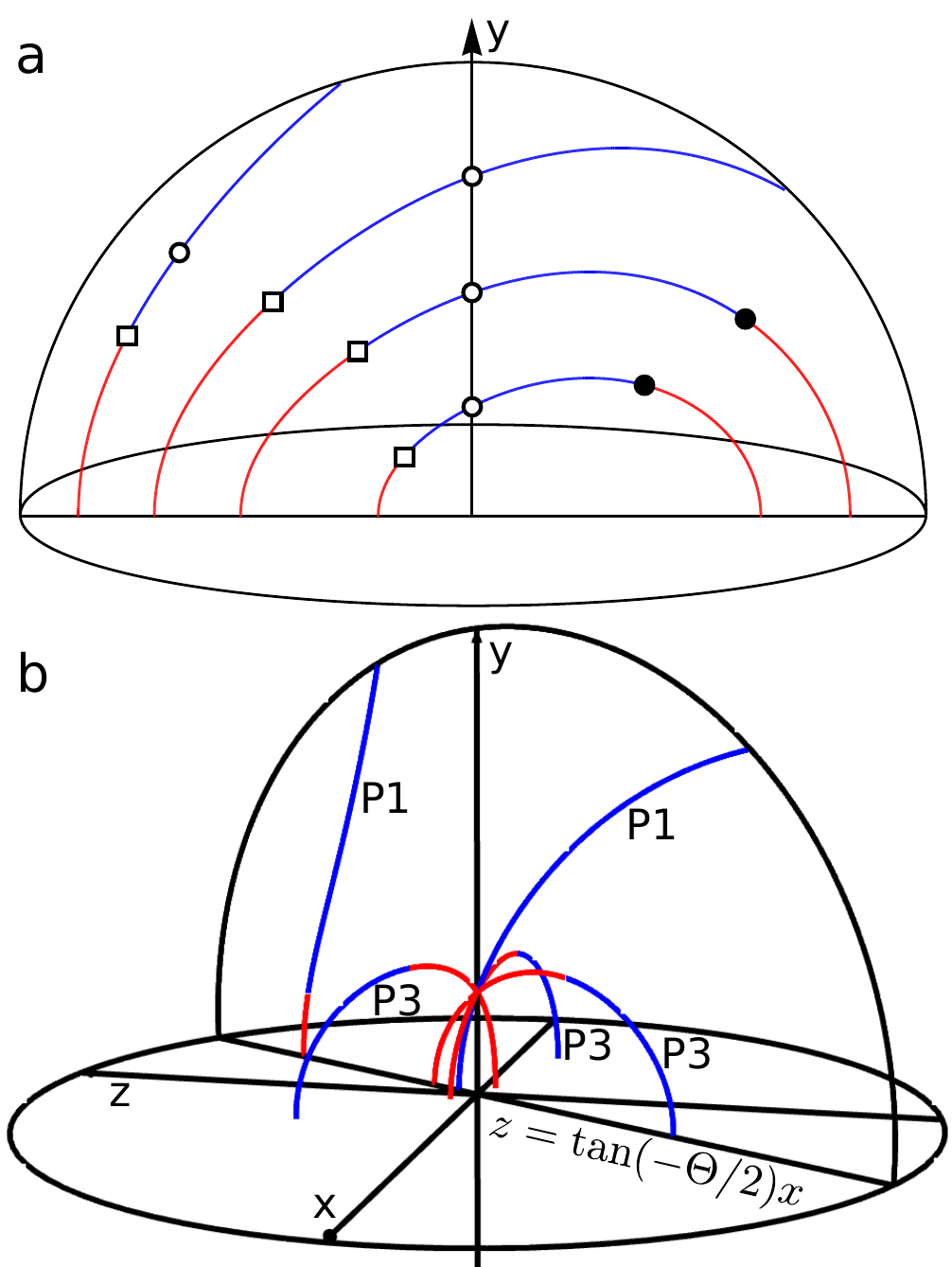}
\caption{Periodic lines in the 3DRPM flow. (a)~Period-1 lines shown in the symmetry plane for $(\Theta,\tau)=(2\pi/3,1.3)$. Elliptic and hyperbolic segments are coloured blue and red respectively. Bifurcation points are illustrated as open squares (period-doubling), open circles (period-tripling) and closed circles (saddle--centre). (b)~The period-1 and period-3 lines for $(\Theta,\tau)=(2\pi/3,1.1\tau_0)$, $\tau_0$ is defined in the text.}
\label{fig:p1_lines_stab}
\end{figure}

The typical nature of period-1 lines in the 3DRPM flow is shown in Fig.~\ref{fig:p1_lines_stab}. We only show points in the $y^+$ hemisphere, since those in the $y^-$ hemisphere can be obtained by reflection through the $xz$-plane. At degenerate points there are three types of bifurcation that occur in the 3DRPM flow: period-tripling bifurcations (open circles), saddle--centre bifurcations (closed circles), and period-doubling bifurcations (squares). In Sections~\ref{sec:period-tripling} and \ref{sec:tangent} we explore their behaviour and implications for transport.

\section{Period-tripling bifurcations} \label{sec:period-tripling}

\subsection{2D model}

\begin{figure}
\includegraphics[width=0.45\textwidth]{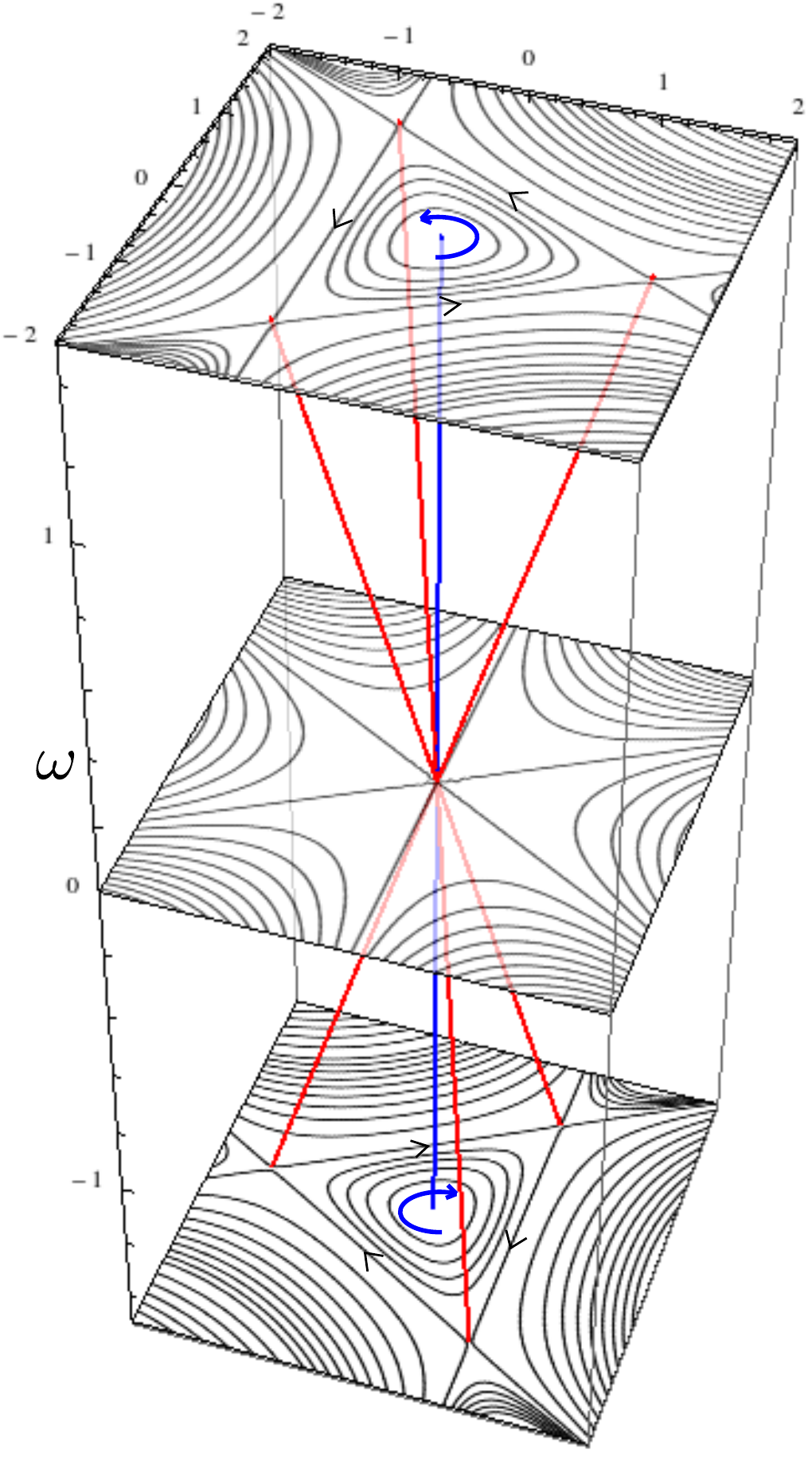}
\caption{A period-tripling bifurcation in the elliptic-umbilic catastrophe, equation~\ref{eq:elliptic_umbilic}, occurs at $\omega=0$. Three hyperbolic fixed points (red) and one elliptic fixed point (blue) coalesce at the bifurcation point.}
\label{fig:twist_schematic}
\end{figure}

Period-tripling bifurcations have been observed in a number of studies\cite{Berry1977,Dullin2000,Mullowney+Julien+Meiss} in 2D and 3D systems, but their transport characteristics in 3D systems have not been studied. In 3D systems these bifurcations occur at points on periodic lines where the rotation angle $\alpha$ (from the eigenvalues $\lambda_{1,2} = e^{\pm i \alpha}$) around an elliptic segment as it is traversed reverses direction, and a chain of period-3 lines intersect at the degenerate point. Due to the complexity of the period-tripling bifurcations in the 3DRPM flow we shall first illustrate the basic structure with a simple model flow derived from an expansion of the flow about a degenerate point with a Poincar\'{e} index of $-2$. This model is a perturbation of the steady 2D six-roll mill flow \cite{Berry1977}, with Hamiltonian
\begin{equation}
F(x,y;\omega) = \frac{x^3}{3} - xy^3 + \omega (x^2 + y^2),
\label{eq:elliptic_umbilic}
\end{equation}
which is related to the elliptic umbilic catastrophe \cite{Thom1989} and the Henon--Heiles potential \cite{Henon1983}. A period-tripling bifurcation occurs when the vorticity $\omega$ is varied as a control parameter, as shown in Fig.~\ref{fig:twist_schematic}. At $\omega=0$ there is no vorticity and there is a degenerate fixed point at the origin, with three stable and three unstable directions, yielding a Poincar\'{e} index of $-2$. For positive or negative $\omega$, the vorticity creates an elliptic fixed point at the origin, and for the Poincar\'{e} index to remain constant three hyperbolic fixed points are created whose heteroclinic manifold connections form the outer barrier for the invariant tori. When transitioning from positive to negative $\omega$, or vice versa, the vorticity changes direction, resulting in a reversal in the orientation of the triangular structure seen in Fig.~\ref{fig:twist_schematic}. A similar phenomenon occurs if vorticity is added to the streamfunction of a $2n$-roll mill for odd $n$. Instead of three stable and unstable directions there will be $n$, and for $\omega\neq 0$ there will be $n$ hyperbolic points arranged in an $n$-gon around an elliptic point, with a Poincar\'{e} index of $1-n$. 

\subsection{3DRPM flow}

\begin{figure*}
\includegraphics[width=0.9\textwidth]{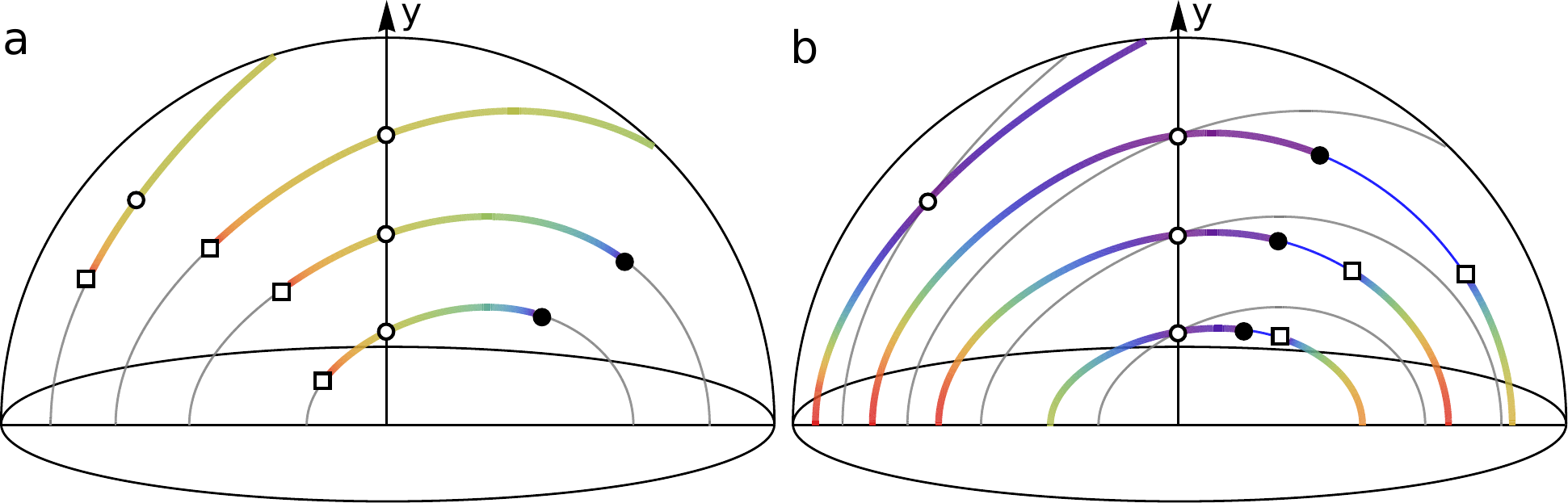}
\caption{(a)~As per Fig.~\ref{fig:p1_lines_stab}a except hyperbolic points are shown as grey and elliptic points are coloured according to local rotation angle $\alpha$, from $0$ (purple) to $\pi$ (red). Points with $\alpha=2\pi/3$ are marked as open circles and correspond to period-tripling bifurcations. (b)~In order to show the intersection of period-1 and period-3 lines in the symmetry plane, the period-1 lines of panel (a) are redrawn in panel (b) as solid gray lines.  The coloured lines of panel (b) are period-3 lines. Elliptic segments are coloured blue and hyperbolic segments are coloured according to the local transverse stretching/contraction factor, with purple (red) corresponding to zero (maximum) stretching/contracting.}
\label{fig:p1_e_p3_h}
\end{figure*}

\begin{figure}
\includegraphics[width=\columnwidth]{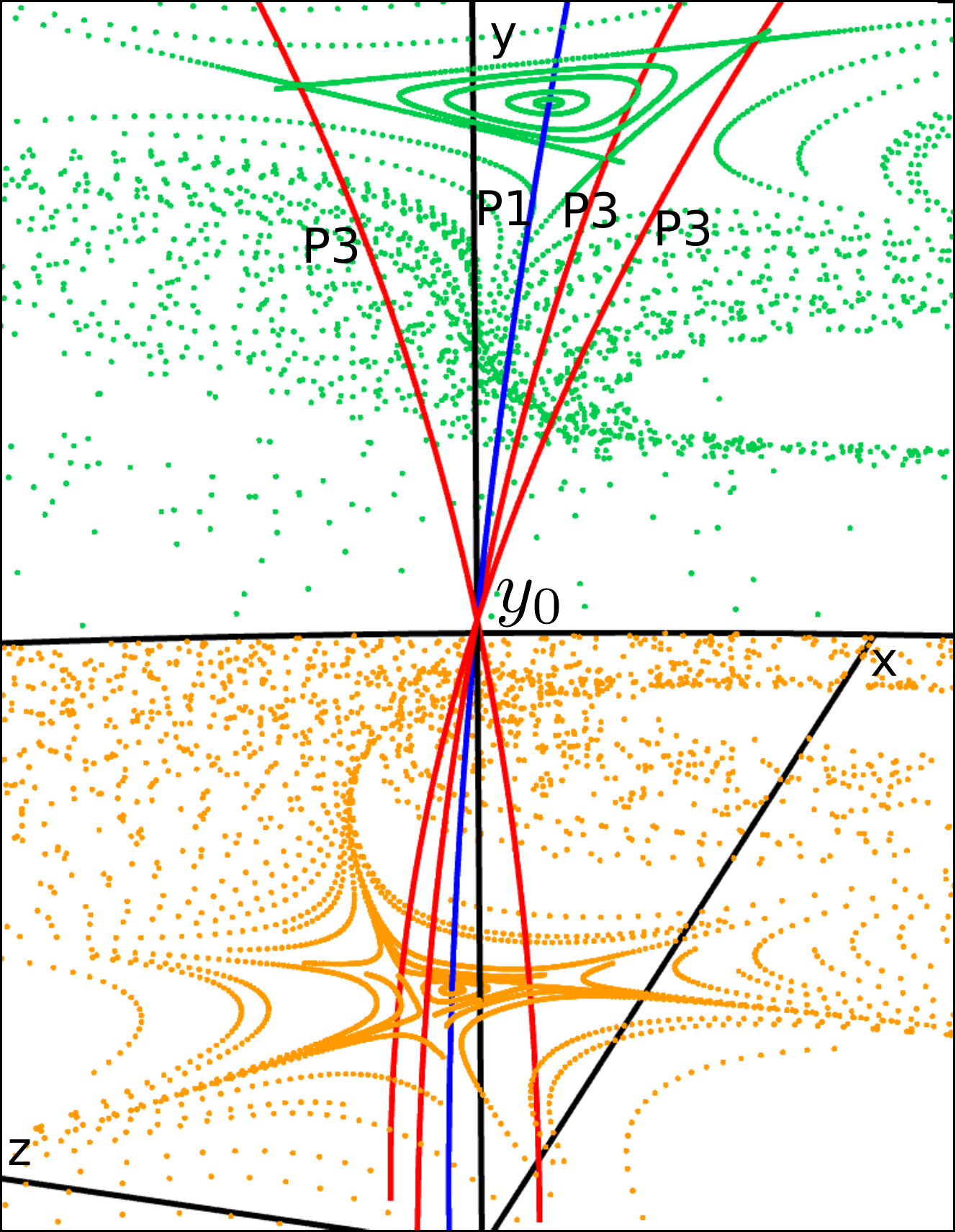}
\caption{Period-tripling bifurcation in the 3DRPM flow with $\tau=1.1\tau_0$. Period-1 and period-3 lines meet at a degenerate point on the $y$-axis (red - hyperbolic, blue - elliptic). Green and orange points correspond to Poincar\'{e} sections at $y$-levels above and below the bifurcation point $y_0$. (Multimedia view) [URL: \url{http://dx.doi.org/10.1063/1.4950763.1}]}
\label{fig:twist_3D}
\end{figure}

Period-tripling bifurcations are present in the 3DRPM flow with rotation angle $\Theta=2\pi/3$ for all values of $\tau$. They manifest at points on periodic lines where the elliptic rotation angle becomes $2\pi/3$, shown in Fig.~\ref{fig:p1_e_p3_h}a as open circles. At these points the map $\hat{Y}_\tau^\Theta$ experiences a $1/3$ resonance, i.e. particles in the vicinity of the elliptic periodic point will approximately return to their initial position after three iterations. If we track particles in the non-rotating (laboratory) frame then the local rotation angle becomes $\epsilon=\alpha-2\pi/3$. The parameter $\epsilon$ varies along the period-1 line, and becomes $0$ at points where $\alpha=2\pi/3$, indicating the presence of degenerate points. At these $1/3$ resonant degenerate points period-tripling bifurcations are generic \cite{Dullin2000}, with the rotation angle $\epsilon$ playing the role of $\omega$ in equation~(\ref{eq:elliptic_umbilic}).

As for the model system given by equation~(\ref{eq:elliptic_umbilic}) there are three hyperbolic lines (period-3 in the rotating dipole frame but period-1 in the non-rotating laboratory frame) that intersect the elliptic period-1 line, resulting in a reversal of triangular invariant tori on each side of the bifurcation point, as per Fig.~\ref{fig:twist_3D} (multimedia view). These invariant tori join to form pyramidal invariant tubes that connect at the bifurcation point. To provide additional evidence of this behaviour, Fig.~\ref{fig:p1_e_p3_h}b shows that period-3 lines (coloured) intersect the period-1 lines (grey) at every location where the rotation angle is $2\pi/3$ (the open circles). Furthermore, the magnitude of stretching along the hyperbolic sections of the period-3 lines is characterised by the magnitude of the logarithm of the transverse eigenvalues $\lambda_{1,2}$ of the Jacobian, which become zero (purple) at the period-tripling bifurcations. This means that although the period-1 rotation angle is not zero, these bifurcation points can be considered as period-3 degenerate points.

\subsection{Creation and annihilation}

\begin{figure*}
  \begin{minipage}[c]{0.65\textwidth}
    \includegraphics[width=\textwidth]{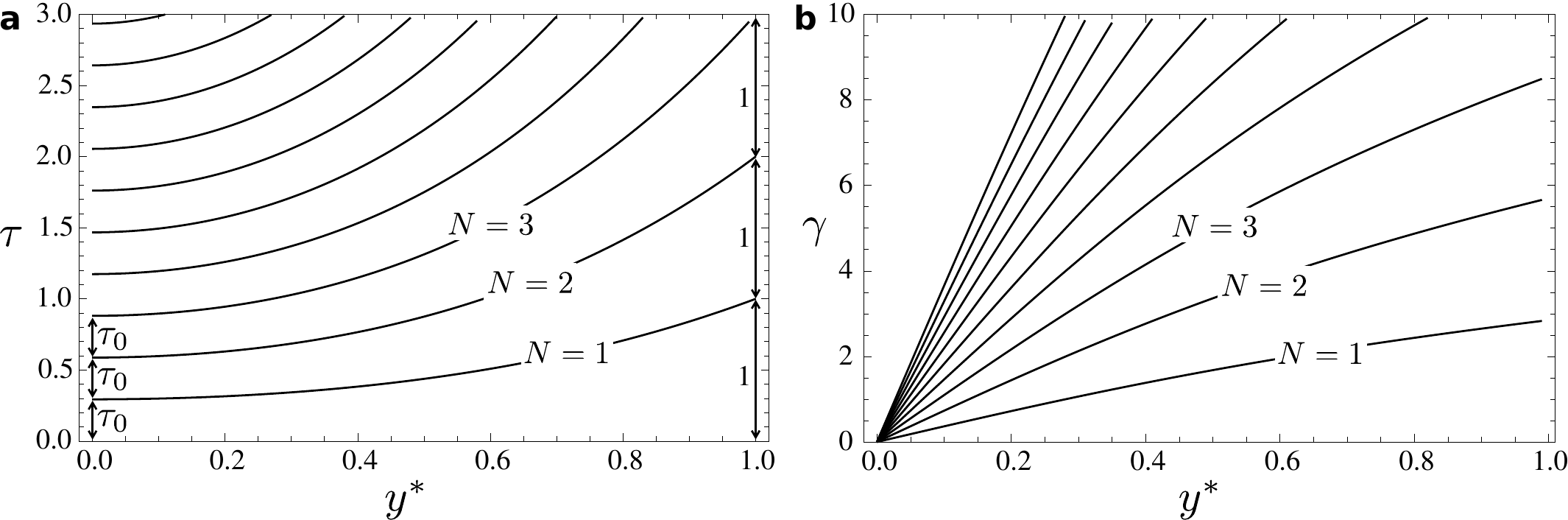}
  \end{minipage}\hfill
  \begin{minipage}[c]{0.3\textwidth}
    \caption{
       (a) The values of $\tau$ for which there is a periodic point on the $y$-axis at $(0,y^*,0)$. $N$ is the number of times the particle is reinjected. (b)~The magnitude of the corresponding shear matrix $D\hat{Y}_\tau$ in eq.~\ref{eq:shear_matrix}.
    } \label{fig:p1_tau_shear}
  \end{minipage}
\end{figure*}


\begin{figure}[b]
\includegraphics[width=0.8\columnwidth]{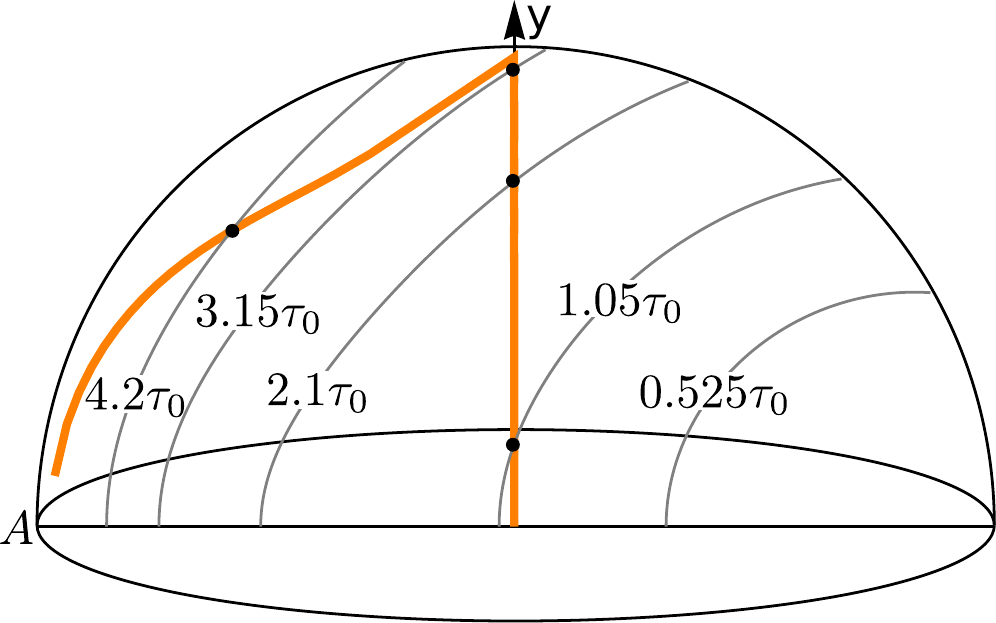}
\caption{The path taken by the period-tripling bifurcation point with increasing $\tau$ (orange) shown with the corresponding period-1 line at various values of $\tau$ (grey). The point where the period-1 lines annihilate is marked as $A$.}
\label{fig:twist_bif_traj}
\end{figure}

\begin{figure}
\includegraphics[width=0.8\columnwidth]{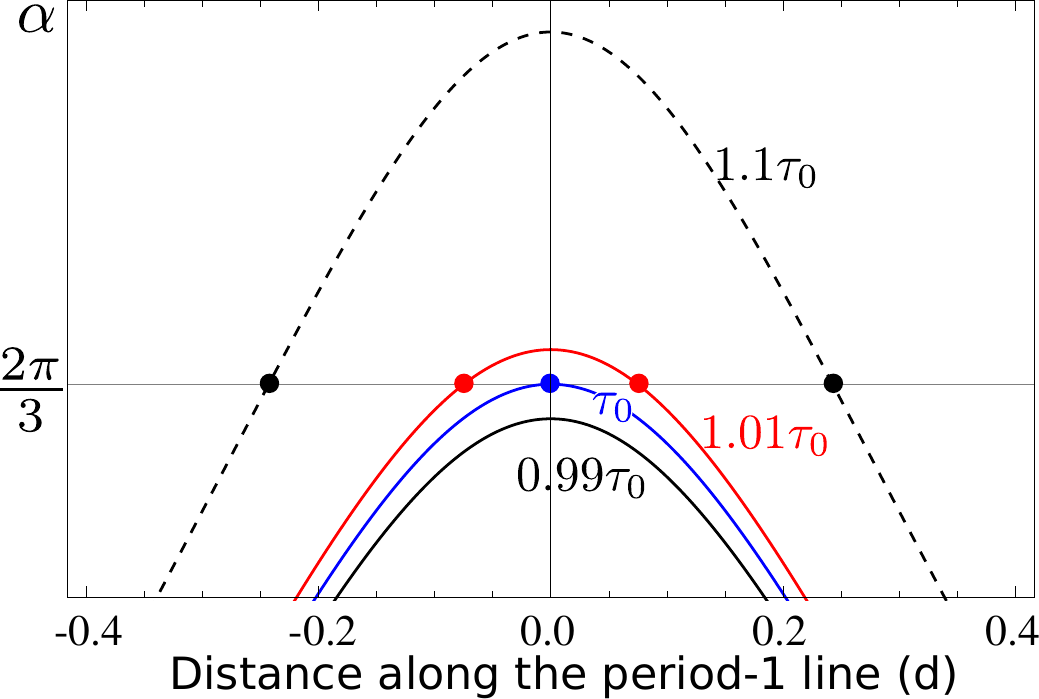}
\caption{The local rotation angle $\alpha$ given by the Jacobian at points a distance $d$ along the period-1 line from the $xz$-plane. Curves are shown for $\tau=0.99\tau_0,\tau_0,1.01\tau_0,1.1\tau_0$. Degenerate points are marked by large dots, where $\alpha=2\pi/3$.}
\label{fig:rotation_angle}
\end{figure}

Considering the creation and annihilation of period-tripling bifurcation points, we consider integer multiples of a critical value of the reorientation period, $\tau_0\approx0.29344$, which corresponds to the return time of a particle initially located at the origin under the steady dipole flow. At the values $\tau=N \tau_0$ the origin is invariant under both the steady dipole advection and rotation that comprise the combined map $Y_\tau^\Theta=R_\Theta^y \hat{Y}_\tau$, hence the origin is a period-1 point, as shown in Fig.~\ref{fig:p1_tau_shear}(a). Moreover, after three iterations of the map $Y_\tau^\Theta$ there is no net local deformation of fluid near the origin, and so this point is a period-3 degenerate point, in particular a period-tripling bifurcation point. In the 3DRPM flow this occurs at any point where a period-1 line intersects the $y$-axis, not just at the origin. For a particle initially located on the $y$-axis to return to its initial position, it must return after the steady dipole advection step, as the rotation cannot move a point that is off the $y$-axis onto the $y$-axis. During the steady dipole advection, fluid at these period-1 points experiences a shear of the form
\begin{equation} \label{eq:shear_matrix}
D\hat{Y}_\tau = \left( \begin{matrix}
1 &0 &0 \\
0 &1 &0 \\
0 &N\gamma_1 &1
\end{matrix} \right)
\end{equation}
where $N$ is the number of times the particle is reinjected during the steady dipole advection, and $\gamma_1$ is the value of the shear experienced during a single reinjection, which depends on the location $y^*$ of the periodic point on the $y$-axis according to Fig.~\ref{fig:p1_tau_shear}(b). Therefore, for a reorientation angle $\Theta$, the Jacobian is given by
\begin{equation}
DY_\tau^\Theta = R_\Theta^y D\hat{Y}_\tau
\end{equation}
with eigenvalues $1,\,\exp(\pm i \Theta)$, and can be diagonalized as $DY_\tau^\Theta = P\mathcal{D}P^{-1}$ where $\mathcal{D}$ is the diagonal matrix
\begin{equation}
\mathcal{D} = \left( \begin{matrix}
1 &0 &0 \\
0 &e^{i\Theta} &0 \\
0 &0 &e^{-i\Theta}
\end{matrix} \right).
\end{equation}
Hence, for $\Theta = 2\pi m/n$, the Jacobian satisfies $(DY_\tau^\Theta)^n = I$, making it a period-$n$ degenerate point. In particular, for $\Theta=2\pi/3$, the periodic points on the $y$-axis are period-3 degenerate points, and thus period-tripling bifurcations.

Tracking the individual period-1 lines with increasing $\tau$, they initially appear as isolated degenerate points resulting from saddle--centre bifurcations in the $xz$-plane, which then form closed loops with a pair of saddle--centre bifurcation points separating elliptic and hyperbolic segments, shown as the solid black circles in Fig.~\ref{fig:p1_lines_stab}a. These loops expand outward as $\tau$ increases, and eventually collide with the spherical boundary, as shown in Fig.~\ref{fig:twist_bif_traj}. Therefore each new period-1 line intersects the $y$-axis at a value $\tau=N\tau_0$, which also corresponds to the creation of a period-tripling bifurcation. At these values of $\tau$, a period-1 line intersects the $y$-axis tangentially, creating a single degenerate point at the origin. This degenerate point is unique because the period-3 lines also intersect the period-1 line and $y$-axis tangentially, meaning there is no reversal in the orientation of the triangular structures. The local rotation still reaches zero, but does not reverse in direction. For a small perturbation $\epsilon$ away from $\tau=N\tau_0$ the tangent intersections become transverse, creating two period-tripling bifurcations that are symmetric about the $xz$-plane, and are orientation reversing. This is depicted in Fig.~\ref{fig:rotation_angle} for $N=1$, showing the rotation angle $\alpha$ as a function of arc-length along the period-1 line starting from the $xz$-plane. Period-tripling bifurcations occur where the rotation angle $\alpha$ reaches $2\pi/3$. For $\tau<\tau_0$ there are no intersections of this period-1 line with the $y$-axis and therefore no period-tripling bifurcations, whereas for $\tau=\tau_0$ there is one intersection and for $\tau>\tau_0$ there are two intersections. The reorientation periods $N\tau_0$ therefore correspond to local flow bifurcations.

\begin{figure}
\includegraphics[width=0.8\columnwidth]{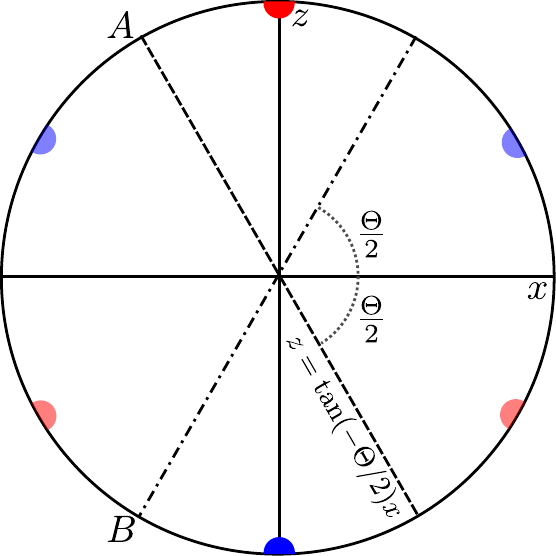}
\caption{The $xz$-plane with the symmetry plane (dashed) and its reflection through the $x$-axis (dot-dashed). The point $A$ corresponds to that of Fig.~\ref{fig:twist_bif_traj}, where period-1 lines annihilate. Dipole position are also shown (red/blue) with the reoriented positions fainter.}
\label{fig:p1_ann_schematic}
\end{figure}

The trajectory of the period-tripling bifurcation point on the first new period-1 line ($N=1$) as $\tau$ is increased is shown in Fig.~\ref{fig:twist_bif_traj} by the orange curve. After appearing at the origin when $\tau=N\tau_0$, it moves up the $y$-axis, to the point where it reaches the spherical boundary. This occurs when $\tau$ is equal to the return time of the streamline on the spherical boundary, which we have scaled to be equal to $1$, multiplied by the number of reinjections $N$, i.e. at $\tau=N$ as shown in Fig.~\ref{fig:p1_tau_shear}(a) by the values of $\tau$ at $y^*=1$. The period-tripling bifurcation point then moves off the $y$-axis, moving towards the $xz$-plane where it is annihilated. This path is traced out for each period-1 line that is created. We find that the period-tripling bifurcation point is annihilated at approximately the same value of $\tau$ as when the period-1 line itself is annihilated, which occurs when the line reaches the point $A$ at the intersection of the $xz$-plane, spherical boundary and symmetry plane, shown in Fig.~\ref{fig:twist_bif_traj}. To find the values of $\tau$ such that the point $A$ is a period-1 point, and hence when the period-1 lines are annihilated, we consider the time $T_{AB}(\Theta)$ it takes for a particle to travel from the point $A$ to its reflection through the $xy$-plane, the point $B$ shown in Fig.~\ref{fig:p1_ann_schematic}. When $\tau = T_{AB}\approx0.9333$, the map $Y_\tau^\Theta$ takes a particle initially located at $A$ to the point $B$ under the steady dipole flow, then the particle is counter-rotated back to $A$, meaning the point $A$ is a period-1 point. Moreover, if the particle is reinjected any number of times through the dipole  but still finishes at $B$ then the point $A$ will still be a period-1 point. As the streamlines on the spherical boundary have the longest return time, which we have scaled to correspond to a value $\tau=1$, $A$ is a period-1 point when $\tau = T_{AB} + j$ where $j$ is the number of times that the particle is reinjected. Combining this with the creation of period-tripling bifurcation points at values $\tau = N\tau_0$, the overall number of period-tripling bifurcation points $\mathcal{N}_{1/3}$ within the 3DRPM flow is a linear function of $\tau$
\begin{align}
\mathcal{N}_{1/3} &\approx \frac{\tau}{\tau_0} -  (\tau - T_{AB}) \nonumber \\
&\approx 2.4 \tau
\end{align}
for $\tau \gg 0$. This shows that the number of period-tripling bifurcation points grows linearly with $\tau$, resulting in more of the stable pyramidal invariant tubes and more of the associated hyperbolic period-3 lines whose manifold intersections can drive chaos. By seeding a large number of particles on a grid in the domain we have seen that the total volume of the invariant tubes decreases as $\tau$ increases even though the total number of invariant tubes increases. It is likely that as $\tau \to \infty$ the total volume of the invariant tubes will approach zero, meaning an approach to global chaos.

\subsection{Impact on transport and manifolds}

\begin{figure}
\includegraphics[width=\columnwidth]{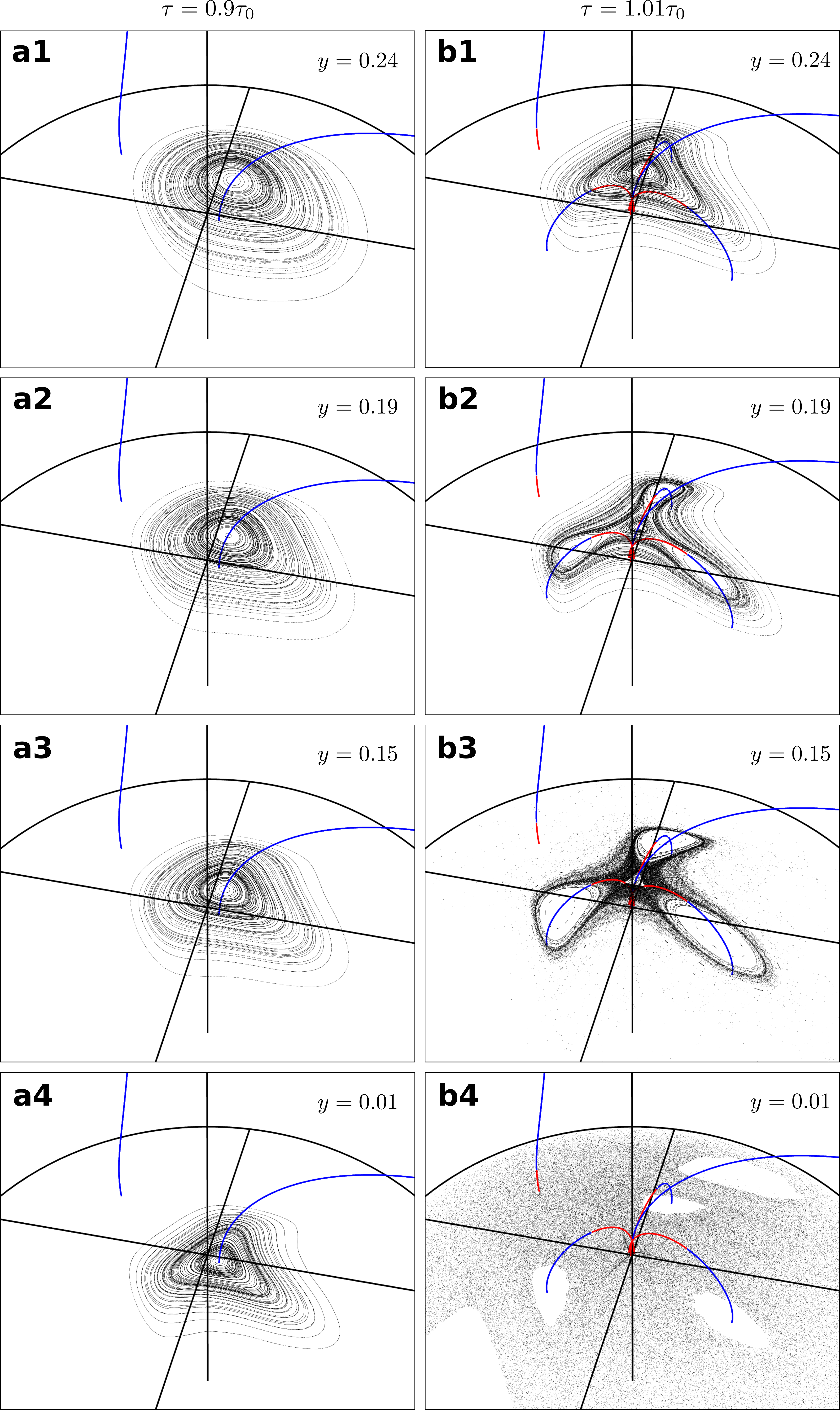}
\caption{Poincar\'{e} sections generated by initially planar (parallel to the $xz$-plane) clusters of particles near the $y$-axis at various values of $y$. Period-1 and period-3 lines are colored according to stability, red - hyperbolic, blue - elliptic. (a1--a4)~$\tau=0.9\tau_0$. There are no period-3 lines, and invariant tori surround the elliptic period-1 line. (b1--b4)~$\tau=1.01\tau_0$. A period-tripling bifurcation occurs on the $y$-axis, that destroys tori, creates sticky regions, creates chaotic regions, alters topology and affects transport in a region of the domain that is vastly more extensive than just the `neighborhood' of the bifurcation point.}
\label{fig:psections}
\end{figure}

\begin{figure*}
  \begin{minipage}[c]{0.65\textwidth}
    \includegraphics[width=\textwidth]{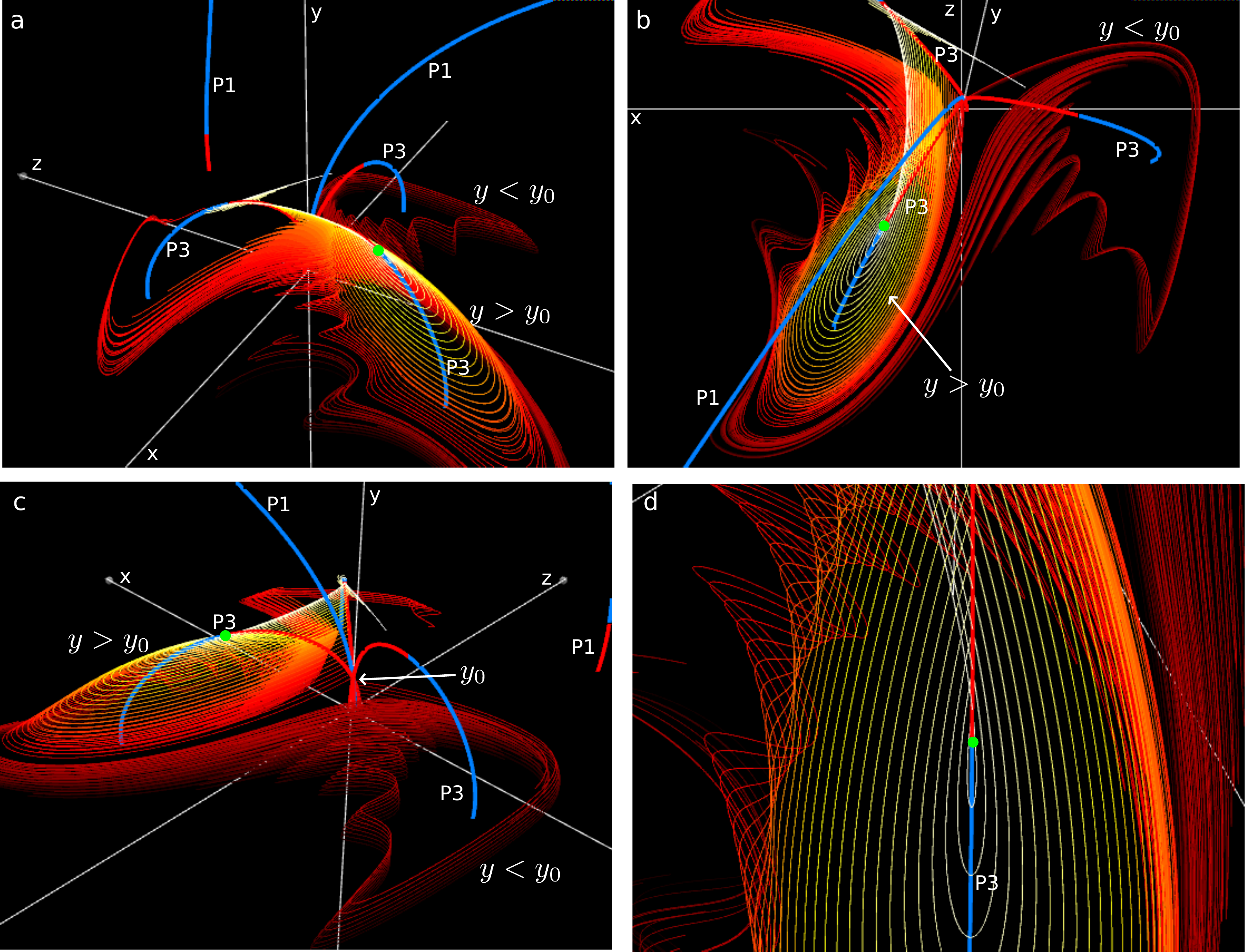}
  \end{minipage}\hfill
  \begin{minipage}[c]{0.3\textwidth}
    \caption{
       (a--d) Four views of the 2D unstable manifold $W^u_{2D}$ for one of the period-3 lines for $\tau=1.1\tau_0$. The manifold consists of the disjoint union of 1D manifolds for points along the hyperbolic segment of the periodic line, colored according to the $y$-coordinate, $y=0$ (dark red) to maximum (white). As in Fig.~\ref{fig:twist_3D} (multimedia view), the point $y_0$ on the $y$-axis -- seen in (c) -- is a period-tripling bifurcation point, and the 2D manifolds for $y<y_0$ and $y>y_0$ form disconnected sheets. The green point marks the location of a tangent saddle--centre bifurcation point, where the stability changes between elliptic and hyperbolic. In (d) it can be seen that close to the bifurcation point the 1D manifolds form parallel homoclinic connections, wrapping around the elliptic line, but at a critical distance away from the bifurcation point waves in the 1D manifolds begin to develop, indicating transverse intersections and chaos. (Multimedia view) [URL: \url{http://dx.doi.org/10.1063/1.4950763.2}]
    } \label{fig:saddle-node_manifolds}
  \end{minipage}
\end{figure*}


The structures associated with period-tripling bifurcations can have a significant influence on the overall transport properties of a fluid flow, as seen in Fig.~\ref{fig:psections}, where the period-tripling bifurcation at $\tau=1.01\tau_0$ (Fig.~\ref{fig:psections}(b)) destroys the invariant tori surrounding the period-1 line that exist at $\tau=0.9\tau_0$ (Fig.~\ref{fig:psections}(a)). The degenerate points associated with the bifurcation have three stable and three unstable directions, creating transport structures similar to hyperbolic points. These stable and unstable directions become the manifolds of the associated period-3 hyperbolic lines away from the bifurcation, as in Fig.~\ref{fig:saddle-node_manifolds}. Also, surrounding the elliptic segments of the period-3 lines there exist invariant tori that join to create impenetrable barriers to particle transport. The locations of the outer-most invariant tori depend on the nature of the manifold intersections, tangential or transverse. If the stable and unstable manifolds intersect tangentially, as is the case at larger $y$-values (closer to white) in Fig.~\ref{fig:saddle-node_manifolds} and also in Fig.~\ref{fig:psections}(b2), then the 1D manifolds form the outermost invariant tori. Descending down the $y$-axis, the manifolds become `wavy', indicating transverse intersections of the stable and unstable manifolds and the existence of a chaotic region. Near the values of $y$ that this first occurs, for example $y=0.19$ as in Fig.~\ref{fig:psections}(b3), particles in the chaotic region near the period-3 invariant tori are loosely trapped in a `sticky' region. Descending further along the $y$-axis (Fig.~\ref{fig:psections}(b4)), the `sticky' region is also destroyed, giving way to wide-spread chaos. Therefore, period-tripling bifurcations can drive global chaos via transverse manifold intersections or create the boundaries for confining regions if the manifold intersections are tangent. 

\begin{figure*}[t]
  \begin{minipage}[c]{0.65\textwidth}
    \includegraphics[width=\textwidth]{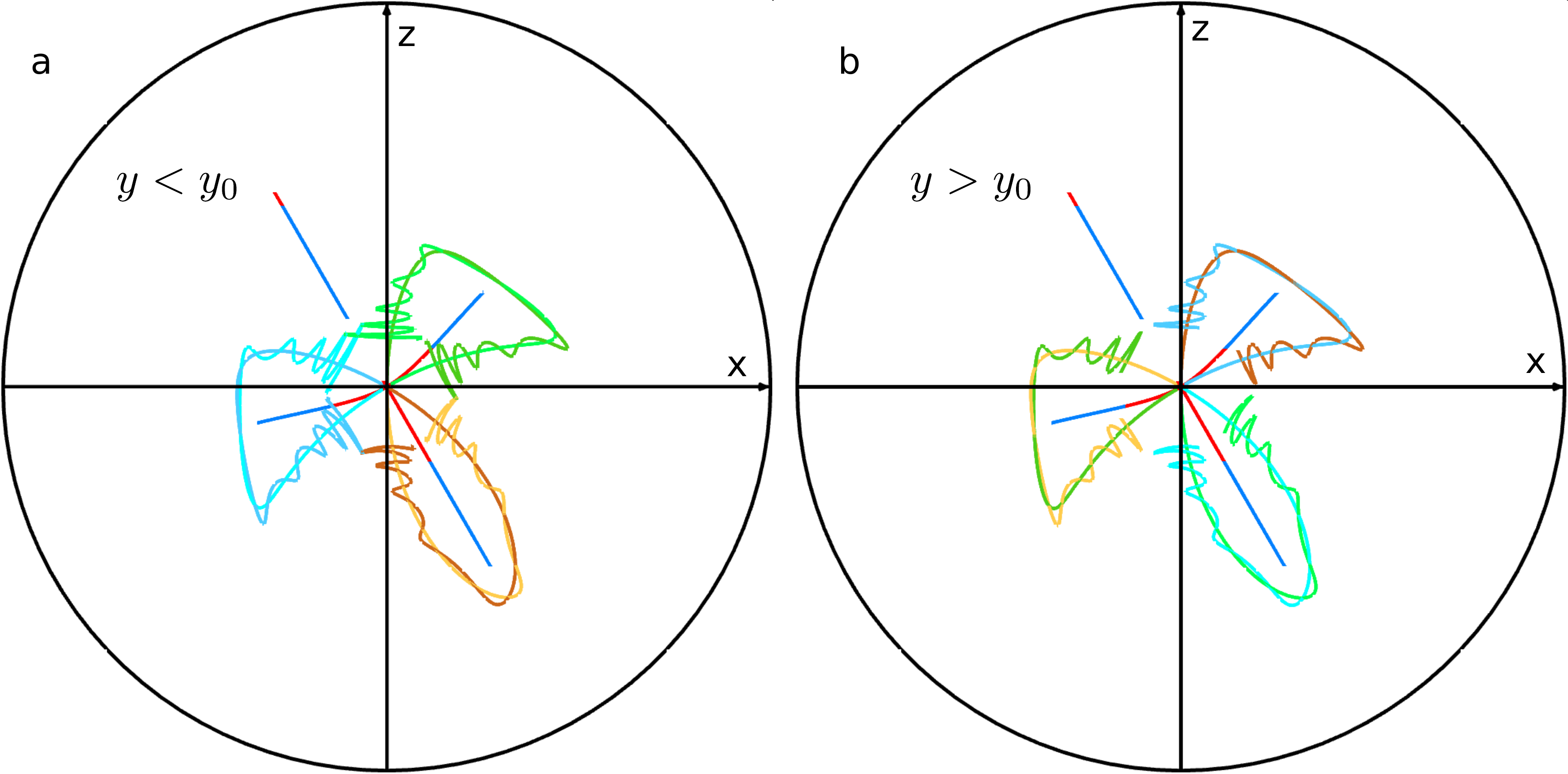}
  \end{minipage}\hfill
  \begin{minipage}[c]{0.3\textwidth}
    \caption{
       Bifurcation of stable/unstable manifolds caused by period-tripling bifurcation, which occurs at $y=y_0$ as in Fig.~\ref{fig:twist_3D} (multimedia view) for $(\Theta,\tau)=(2\pi/3,1.01\tau_0)$. Manifold pairs associated with each period-3 line are shown as different shades of the same colour, e.g. green and dark green. (a)~$y<y_0$, manifolds form homoclinic connections. (b)~$y>y_0$, manifolds form heteroclinic connections.
    } \label{fig:mfold_bif}
  \end{minipage}
\end{figure*}


As well as a bifurcation in local stability, period-tripling bifurcations also create a bifurcation in the manifolds associated with the hyperbolic period-3 lines, as demonstrated in Fig.~\ref{fig:mfold_bif}. The stable and unstable manifolds for single points on each of the period-3 lines are shown at values $y<y_0$ (left panel) and $y>y_0$ (right), where $y_0$ is the bifurcation point. The manifold pairs $W^{s,u}$ associated with each period-3 line are coloured with two shades of the same colour (e.g. stable - green, unstable - dark green etc.). We see that for $y<y_0$ (left) the manifolds form transverse homoclinic connections since the same colours intersect, whereas for $y>y_0$ (right) the connections become heteroclinic, and so while the orientation of the triangular island structures are reflected across the bifurcation point, the global arrangement of structures remains essentially the same. This is only possible if there is a change between heteroclinic and homoclinic connections across the bifurcation point. Considering the entire 2D unstable manifold $W^s_{2D}$ associated with only one of the period-3 lines, as in Fig.~\ref{fig:saddle-node_manifolds} (multimedia view), there is a disconnection of the manifold sheet as the period-3 point crosses the bifurcation point. The bifurcation point can either be thought of as a point of discontinuity for the manifold sheet, or the two segments of the period-3 line, separated by the bifurcation point, can be considered as separate entities with their own manifold structures.

Therefore, the period-tripling bifurcation points themselves do not have a significant impact on global transport, as they occur as an isolated unstable point along an otherwise elliptic periodic line, but they organize vast transport structures that generate chaos and form barriers to transport. 

For other values $\Theta=2\pi m/n$ with $n$ odd we see similar behaviour. Rather than period-tripling bifurcations they are $n$-tupling. Instead of three period-3 hyperbolic lines there are $n$ period-$n$ lines and the degenerate point has a Poincar\'{e} index of $1-n$. These cases are analogous to the $2n$-roll mill, with the arc-length along the period-1 line acting as the control parameter $\omega$.

\section{Tangent bifurcations} \label{sec:tangent}

\begin{figure}
\includegraphics[width=0.8\columnwidth]{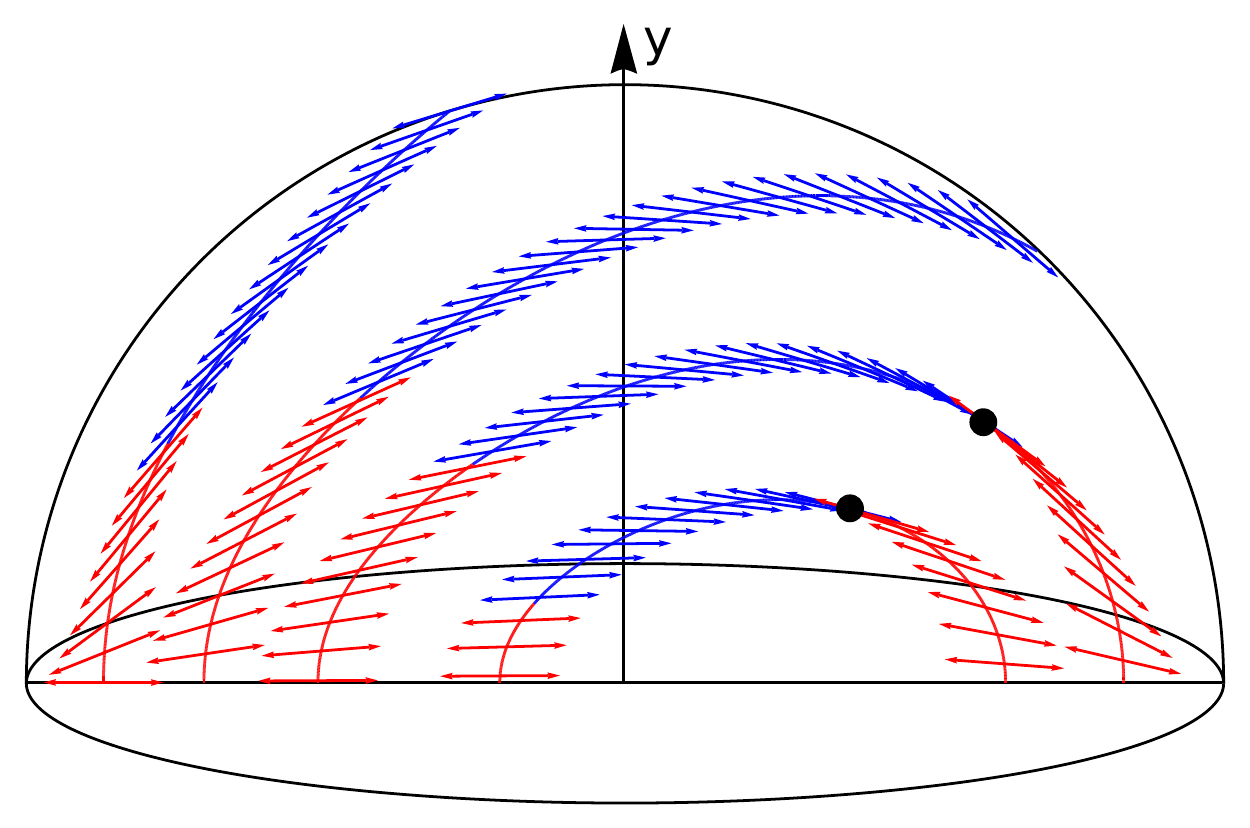}
\caption{The period-1 lines from Fig.~\ref{fig:p1_lines_stab} are reproduced with one of the non-null eigenvectors also shown. The other non-null eigenvector is always normal to the symmetry plane (into the page) as a result of the reflection reversal symmetry - equation~(\ref{eq:ref-rev_sym}). Points where the null direction (tangent of the lines) becomes tangent with the other eigenvectors are shown as black circles. These correspond to tangent saddle--centre bifurcations.}
\label{fig:p1_evectors}
\end{figure}

\begin{figure*}
  \begin{minipage}[c]{0.6\textwidth}
    \includegraphics[width=\textwidth]{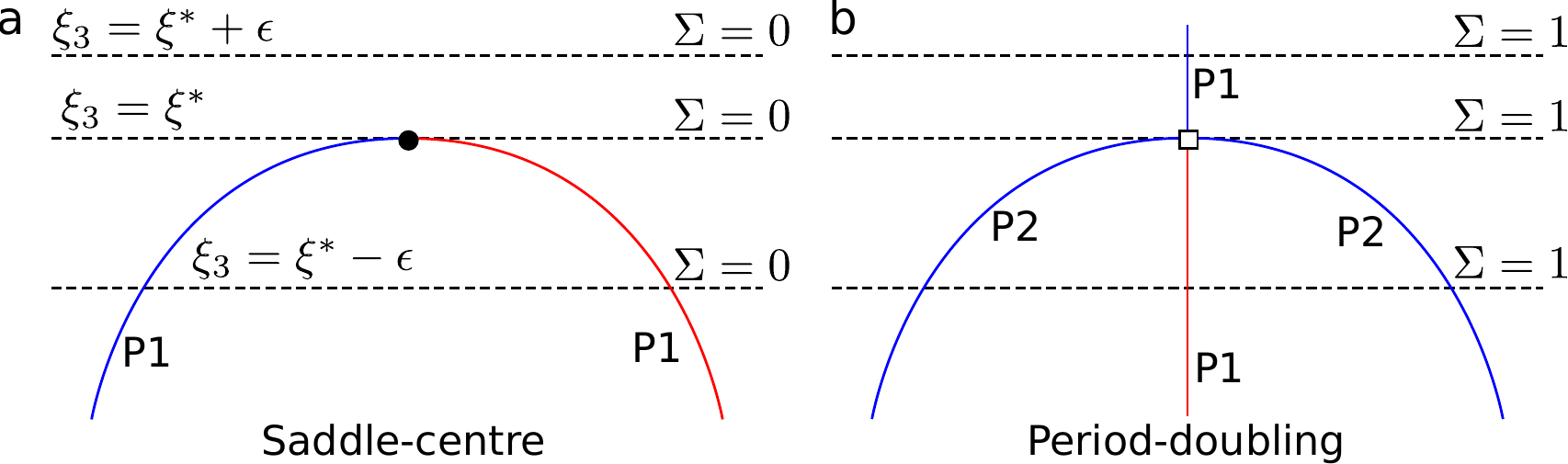}
  \end{minipage}\hfill
  \begin{minipage}[c]{0.35\textwidth}
    \caption{
       Sketch of the periodic lines for two types of tangent bifurcation: (a) saddle--centre (b) period-doubling. In each case the elliptic segments are coloured blue and contribute $+1$ to the Poincar\'{e} index $\Sigma$, and hyperbolic segments are coloured red and contribute $-1$ to $\Sigma$. The local/global invariant $\xi_3$ from equation~(\ref{eq:ess_2d}) acts as the control parameter, with the bifurcations occurring at $\xi_3=\xi^*$. Conservation of the Poincar\'{e} index $\Sigma$ for each locally 2D system $\xi_3=c$ constrains the possible types of tangent bifurcation.
    } \label{fig:tangent_schematic}
  \end{minipage}
\end{figure*}

Tangent bifurcations occur in 3D systems when the null direction ($\bm{v}_3$) of the Jacobian becomes tangent to the plane spanned by the other two eigenvectors ($\bm{v}_{1,2}$). Formally a point $\bm{x}_0$ is a tangent bifurcation point if the vectors $\bm{w}_{1,2,3}$ are linearly dependent, where
\begin{equation}
\bm{w}_i = \lim_{\substack{\bm{x} \to \bm{x}_0 \\ \bm{x} \in \text{P1 line}}} \bm{v}_i (\bm{x}).
\end{equation} 
There are infinitely many possibilities for the different types of tangent bifurcations, based on the periodicity of the periodic line that becomes tangent, for instance saddle--centre bifurcations occur when a period-1 line becomes tangent and period-doubling bifurcations occur when a period-2 line becomes tangent\cite{Note2}. By plotting the eigenvectors with the periodic line, tangent bifurcations can be identified as points where the span of the non-null eigenvectors becomes tangent to the periodic line, as per Fig.~\ref{fig:p1_evectors}. Alternatively, in systems which admit a global invariant, tangent bifurcations occur at points where periodic lines become tangent to an invariant surface. When the null direction becomes tangent to an invariant surface the three eigenvectors are linearly dependent, and hence no longer distinct. For incompressible flows the product of the eigenvalues is always equal to $1$, so there are only two possibilities for the eigenvalues: $\lambda_{1,2} = \pm 1$ and $\lambda_3=1$. This means that these tangent points are necessarily degenerate, resulting in bifurcations of local stability. This provides a simple diagnostic to determine the locations of some degenerate points in systems with an invariant. In systems that do not admit a global invariant, such as the 3DRPM flow, the flow becomes essentially 2D near the periodic line, according to equation~(\ref{eq:ess_2d}), and so the variable $\xi_3$ is a local invariant. Like global invariants, tangent bifurcations occur where a periodic line becomes tangent to isosurfaces of the local invariant.


Conservation of the Poincar\'{e} index constrains the possible types of bifurcation that can occur via tangent bifurcations. Two of the possibilities are illustrated in Fig.~\ref{fig:tangent_schematic}: a saddle--centre bifurcation (left) and a period doubling bifurcation (right), both of which occur in the 3DRPM flow (Fig.~\ref{fig:p1_lines_stab}). Note that saddle--centre bifurcations are sometimes also referred to as tangent bifurcations in the context of 2D systems, but here we use the term to refer to the broader class of bifurcations for 3D systems that includes saddle--centre bifurcations. In each case the bifurcation point must occur where the periodic lines become tangent to the invariant surfaces $\xi_3=c$, otherwise the Poincar\'{e} index $\Sigma$ would not be conserved. These constraints allow us to make {\em a priori} deductions with limited information, for example:
\begin{itemize}
\item If there is a point on a periodic line that lies tangent to an invariant surface, and no other periodic lines intersect at the same point, then the tangent point must be a saddle--centre bifurcation point separating elliptic and hyperbolic segments, as per Fig.~\ref{fig:tangent_schematic}a.
\item Additionally, if there exists a saddle--centre bifurcation point on a periodic line, then the tangent to the periodic line at that point forms one of the tangent vectors of the local/global invariant.
\item If there is a point on a periodic line that is tangent to invariant surfaces, and the periodic line has the same stability on each side of the tangent point e.g. both elliptic, then there must exist another periodic line (possibly of different periodicity) that also intersects at the tangent point. This is the case for the period-2 line in the period-doubling bifurcation, as in Fig.~\ref{fig:tangent_schematic}b.
\end{itemize}

Considering the impact that tangent bifurcations have on transport, we primarily focus on saddle--centre bifurcations as they provide a complete picture for the bifurcation sequence that occurs near period-tripling bifurcations in the 3DRPM flow. At a critical value $y>y_0$, the period-3 lines associated with the period-tripling bifurcation undergo saddle--centre bifurcations, dividing them into elliptic and hyperbolic segments, as seen in Figs.~\ref{fig:p1_e_p3_h}b, \ref{fig:saddle-node_manifolds} (multimedia view). While not the focus of this study, period-doubling bifurcations are equally important for transport. Cascades of period-doubling bifurcations are a common route to chaos in 2D systems \cite{Feigenbaum1979} and we expect similar behaviour for 3D systems, though the chaos may be restricted to approximately 2D structures.

Saddle--centre bifurcations are commonly found in 2D conservative systems, resulting in the creation of a pair of periodic points, one elliptic and one hyperbolic. For 3D conservative systems the third dimension can act as the control parameter for essentially 2D transport. Thus saddle--centre bifurcations in 3D conservative systems create elliptic and hyperbolic segments of periodic lines. Enclosing the elliptic segment is an invariant tube, yielding an isolated non-mixing region. At the saddle--centre bifurcation point the elliptic segment and hence the invariant tube converges to a point, creating a `cap' for the tube. For the 3DRPM flow the cap is formed by the tangent connections of the 2D stable and unstable manifolds $W^{s,u}_{2D}$, as seen in Fig.~\ref{fig:saddle-node_manifolds} (multimedia view). However at a critical distance away from the bifurcation point the 2D manifolds intersect transversally, indicated by the `wavy' pattern that appears at smaller $y$ values (closer to red) in Fig.~\ref{fig:saddle-node_manifolds}d (multimedia view). This transverse intersection means the 2D manifolds no longer form the outer boundary of the invariant tube, but rather there is a bounding ergodic region. We expect this phenomenon to be generic, as the distance from the bifurcation point in the transverse direction $\xi_3$ to the essentially 2D transport can act as a perturbation parameter.

Therefore the framework of tangent bifurcations, in particular the restrictions imposed by conservation of the Poincar\'{e} index, provide a simple diagnostic tool for the analysis of periodic lines. For each periodicity there exists at least one distinct type of tangent bifurcation, e.g. saddle--centre bifurcations occur when a period-1 line becomes tangent, and period-doubling bifurcations occur when a period-2 line becomes tangent. These different types of tangent bifurcations can have vastly different impacts on transport. For instance, saddle--centre bifurcations yield both isolated non-mixing regions and the possibility of locally chaotic regions, whereas period-doubling bifurcations can create regions of chaos via period-doubling cascades, though possibly only two-dimensional chaos.

\section{Conclusions}

Degenerate points play a more important role than generally supposed in the organization of transport structures, representing bifurcations in local stability and transport topology. Typically in 2D systems there exists a perturbation parameter that controls bifurcations, but in 3D the extra dimension may act as the perturbation parameter for locally 2D transport. In 3D systems with periodic lines rather than isolated periodic points, the null direction associated with the periodic line acts as a local invariant producing bifurcations in the essentially 2D transport.

We have studied the bifurcations that occur in a 3D model fluid flow, the 3DRPM flow, which are generic to a wide range of 3D conservative systems. Period-tripling bifurcations result in a local reversal of some coherent structures and are generic to $1/3$ resonances, i.e. when the local elliptic rotation angle $\alpha$ is $2\pi/3$. In the 3DRPM flow with reorientation angle $\Theta=2\pi/3$ these bifurcations occur at every intersection of period-1 lines with the $y$-axis, and for reorientation angles of the form $2\pi m/n$ with $n$ odd similar bifurcations occur. Rather than period-tripling they are $n$-tupling bifurcations, occurring at the intersections of $n$ period-$n$ hyperbolic periodic lines and an elliptic period-1 line. Even though each period-tripling bifurcation results in a single unstable point on an elliptic periodic line, it also controls the associated period-3 hyperbolic lines whose manifolds play a significant role in transport organization, destroying tori, creating `sticky' regions and creating wide-spread chaos. Since the nature of the manifold intersections, either tangential or transverse, can depend on the position along the periodic lines, combinations of chaotic, sticky, and confining regions can occur in a single 3D flow, leading to complex 3D transport.

Also observed in the 3DRPM flow are saddle--centre and period-doubling bifurcations; the former creating barriers to transport and the possibility of chaos, and the latter being a common route to chaos in 2D systems. Both of these types of bifurcation can be categorized as tangent bifurcations, which occur when periodic lines are tangent to local/global invariant surfaces. These are particularly easy to detect in systems with global invariants, and can also be detected by considering the eigenvectors associated with the periodic line when there is no global invariant, such as in the 3DRPM flow. The restrictions imposed by conservation of the Poincar\'{e} index place constraints on the types of possible tangent bifurcations, and can also be used to determine properties such as stability and the existence of higher or lower order periodic points.

The period-tripling bifurcations have also been studied in 2D systems in the context of so-called `twistless-tori'\cite{Dullin2000}, i.e. invariant tori surrounding elliptic periodic points where the rotation number does not vary monotonically. These twistless-tori produce a number of interesting bifurcation phenomena that are not predicted by the KAM-theorem, such as reconnection bifurcations where chains of periodic points merge and annihilate each other. Dullin \emph{et al.}\cite{Dullin2000} have shown that twistless-tori generically appear as a result of period-tripling bifurcations in 2D systems, and our recent results suggest that similar phenomena are observed in 3D systems featuring period-tripling bifurcations, such as the 3DRPM flow. This will be the subject of a future publication.

\begin{acknowledgments}
L. Smith is funded by a Monash Graduate Scholarship and a CSIRO Top-up Scholarship.
\end{acknowledgments}

\appendix

\end{document}